\documentclass[12pt,a4paper]{article}
\usepackage{amsmath,amssymb,amsthm,amsfonts}
\usepackage[margin=1.0in,rmargin=1.in,tmargin=2cm,bmargin=3cm]{geometry}
\usepackage{cite}
\usepackage{graphicx}
\usepackage{subcaption}
\usepackage{epsfig}
\usepackage{hhline}
\usepackage[pagebackref=false]{hyperref}
\usepackage[affil-it]{authblk}
\numberwithin{equation}{section}

\def\e{\epsilon}

\def\l{\lambda}

\def\be{\begin{equation}}
\def\ee{\end{equation}}
\def\bea{\begin{eqnarray}}
\def\eea{\end{eqnarray}}
\def\bg{\begin{align}}
\def\eg{\end{align}}

\def\tr{{\rm tr}}

\makeatletter
\renewcommand\section{\@startsection {section}{1}{\z@}%
	{-3.5ex \@plus -1ex \@minus -.2ex}
	{2.3ex \@plus.2ex}%
	{\normalfont\large\bfseries}}
\renewcommand\subsection{\@startsection{subsection}{2}{\z@}%
	{-3.25ex\@plus -1ex \@minus -.2ex}%
	{1.5ex \@plus .2ex}%
	{\normalfont\bfseries}}
\makeatother

\begin{document}
	\begin{titlepage}
		{\title{{More on Phase Transition and R\'enyi Entropy}}}
		\vspace{.5cm}
		\author{Saeed Qolibikloo \thanks{s.qolibikloo@mail.um.ac.ir}}
		\author{Ahmad Ghodsi \thanks{a-ghodsi@ferdowsi.um.ac.ir}}
		\vspace{.5cm}
		\affil{ Department of Physics, Faculty of Science,     
			\hspace{5.5cm}Ferdowsi University of Mashhad, 	
			\hspace{7.5cm} Mashhad, Iran}
		\renewcommand\Authands{ and }
		\maketitle
		\vspace{-12cm}
		\begin{flushright}
		\end{flushright}
		\vspace{10cm}
		\begin{abstract}
In this paper, we study the scalar field condensation around the hyperbolic black hole solutions in the Einstein and Gauss-Bonnet gravities. We investigate the R\'enyi entropy and inequalities governing on it under this phase transition. Our numerical computations show that for the positive values of the Gauss-Bonnet coupling and below a critical temperature one of these inequalities is violated. This puts more restrictions on the allowed values of the Gauss-Bonnet coupling.
		\end{abstract}
	\end{titlepage}
\section{Introduction}\label{sec.2.1}
The study of Entanglement Entropy (EE) has been a significant subject in recent years. As an evidence to its importance, one could point to its many applications in quantum information theory and quantum computation \cite{Abramsky:2003}, condensed matter physics \cite{Calabrese:2004eu,Hamma:2005,Calabrese:2005zw,Kitaev:2005dm}, quantum gravity and specially in holography \cite{Ryu:2006bv,Ryu:2006ef,Nishioka:2009un,VanRaamsdonk:2009ar,VanRaamsdonk:2010pw,Takayanagi:2012kg,Bianchi:2012ev, Myers:2013lva,Balasubramanian:2013rqa}. 

Generally speaking, when a quantum system is in a pure state, EE is a good measure for the degree of entanglement between any two subsystems. If the composite system is in a mixed state, other metrics such as the mutual information have to be used. To define EE more carefully, consider a system in a pure quantum state, composed of two subsystems $A$ and its complement $A^{c}$, whose state can be described by a density matrix $\rho$. The entangling surface $\partial A$ is defined as a boundary surface of the spatial region $A$ in the quantum field theory (QFT) under consideration.

By tracing over the degrees of freedom in $A^{c}$, one can construct a reduced density matrix $\rho_A=\tr_{A^c} \rho$. The EE between these two subsystems is measured by the Von Neumann entropy of the reduced density matrix $i.e.$ $S_{EE}=-\tr \rho_{A} \log \rho_{A}$. The computation of  $\log \rho_{A}$ is a hard task even in the simplest cases of two dimensional QFTs. Instead, one can compute $\tr\rho^{n}_{A}$ and use it for another measure, known as Entanglement R\'enyi Entropy (ERE) \cite{Renyi0}
\begin{align}
S_n=\frac{1}{(1-n)} \log \tr \rho^{n}_{A}.
\end{align}
Taking the limit $n \to 1$ one yields the entanglement entropy, $i.e.$ $S_{EE}=\lim_{n \to 1} S_n$. 

ERE contains more useful information about the spectrum of $\rho_A$ than EE. Indeed, one can recover the whole spectrum of the reduced density matrix $\rho_A$ only by knowing the R\'enyi entropies for all integers $n>0$ \cite{Li-Haldane}. For example, $S_{\infty}=-\log \lambda_{max}$, where $\lambda_{max}$ is the largest eigenvalue of $\rho_A$ or $S_0=\log D$, in which, $D$ is the number of non-vanishing eigenvalues of $\rho_A$. In particular, $S_2=-\log \tr \rho^{2}_{A}=-\ln P$, where $P$ is the probability of finding two systems in the same state, after the measurement in a diagonalized basis. These special cases are called the min-entropy, the Hartley or the max-entropy and the collision entropy respectively.

As we mentioned, the computation of $\tr \rho_A^n$ is much easier than the computation of $\tr \log \rho_A$. In the context of Conformal Field Theories (CFT)s, a standard way to calculate $\tr \rho_A^n$ (and consequently $S_{EE}$) is  the ``Replica Trick". In this method, after a Wick rotation from the flat Minkowski metric where the CFT lives to a metric with Euclidean signature, the desired $\tr \rho_A^n$ operator is given in terms of the path integral on an $n$-sheeted Riemann surface
\begin{align}
\tr \rho_A^n=\frac{Z_n}{(Z)^n}\,,
\end{align}
where $Z$ is the partition function of the original space-time and $Z_n$ is the partition function on a singular space, which is constructed by gluing $n$ copies of the original space along the boundary $\partial A$. At the end, $\tr \rho_A^n$ essentially becomes product of the two-point correlation functions of the twisted vertex operators \cite{Calabrese:2004eu,Cardy:2007mb}  which is very hard to calculate, except in very specific cases, for example see \cite{Nishioka:2006gr, Velytsky:2008rs, Buividovich:2008kq} or \cite{Nishioka:2009un} and references therein. In any case, after the computation of $\tr \rho_A^n$, we can substitute it in the following equation to compute $S_{EE}$,
\begin{align}
S_{EE}=-\frac{\partial}{\partial n}\log \tr \rho_A^n\Big|_{n=1}\, .
\end{align}
In the context of holography, EE  first appeared in \cite{Ryu:2006bv} where the authors, Ryu and Takayanagi (RT), gave a simple yet ingenious way for the computation of the $S_{EE}$. In their method, the EE between the two spatial regions $A$ and its complement in the $d$-dimensional boundary CFT, is found to be proportional to the extremum  of the surface area of the bulk hypersurfaces $\Sigma$, which are homologous to the region $A$, $i.e.$ $\partial A=\partial \Sigma$.
\begin{align}
S_{EE}(A)=\mathrm{Ext}\left[\frac{\mathcal{A}(\Sigma)}{4 G_N}\right]\,.
\end{align}

The resemblance between the formula above and the Bekenstein-Hawking formula for black hole (BH) thermal entropy $S_{BH}=\frac{A}{4 G}$, is extraordinary. However, the surface $\Sigma$ does not need to coincide with any BH event horizon in general. This striking similarity prompted physicists to look for a derivation of the RT conjecture, for example, see \cite{Lewkowycz:2013nqa}. 

In paper \cite{Casini:2011kv}, authors give a novel method for the computation of the entanglement entropy of a Spherical Entangling  Surfaces (SES) in a CFT, on a flat Minkowski space-time. 
In what follows we present a quick review of their approach.

First, let us consider a quantum system in a $d$-dimensional flat Minkowski space-time $R^{1,d-1}$, which is divided into two subsystems, $A$ and its complement $A^c$. The subsystem $A$ consists of a spatial region inside a ball of radius $R$, whose boundary (at time $t=0$) is denoted by $\partial A$. The subsystem $A$ can be described by a reduced density matrix $\rho_A=\tr _{A^c}\rho$, where $\rho$ is the density matrix of the vacuum state of the whole system. The Cauchy development $D(A)$, by definition is the set of all space-time events $p\in R^{1,d-1}$, through which, every non-space-like curve intersects with the region $A$ at least once.
After a Wick rotation the metric is 
\begin{align}
ds^2_{R^d}=dt^2_E+dr^2+r^2 d\Omega^2_{d-2}\,,
\end{align}
where $t_E$ is the Euclidean time, $r$ is the radial coordinate on the constant time slice $(t_E=0)$ and $d\Omega^2_{d-2}$ is the metric on a unit $(d-2)$-sphere. Notice that, SES in this metric is the region $(t_E,r)=(0,R)$.
After a series of conformal transformations (first $z=r+it_E$, second $\exp(-w)=\frac{R-z}{R+z}$ and finally $w=u+\frac{i\tau_E}{R}$) the metric can be written as
\begin{align}
ds^2_{S^1\times H^{d-1}}=\Omega^2 ds^2_{R^d}=d\tau^2_E+R^2(du^2+\sinh^2u d\Omega^2_{d-2})=d\tau^2_E+R^2d\Sigma^2_{d-1} \,,
\end{align}
where $d\Sigma^2_{d-1}$ is the metric on a unit $(d-1)$-dimensional hyperbolic plane and $\Omega$ is a conformal factor $\Omega=\frac{2R^2}{|R^2-z^2|}=\left|1+\cosh w\right|$.

In short, $D(A)$ is conformally mapped to $S^1\times H^{d-1}$. In other words after the inverse Wick rotation, we will see that the causal development of the ball enclosed by the entangling surface (SES) is conformally mapped into a hyperbolic cylinder $R\times H^{d-1}$, or equivalently the minimal surface of the RT conjecture gets mapped to the horizon of the topological BH. This mapping also translates the vacuum of the CFT in the original Minkowski space-time into a thermal bath with the temperature $T_0=\frac{1}{2\pi R}$ in the hyperbolic cylinder. 

Note that, the curvature scale on the hyperbolic spatial slice, is equal to the radius $R$ of the original SES. Just like any other operator in a CFT, the density matrix in the new space-time $R^1\times H^{d-1}$, can be achieved by a unitarity transformation of the density matrix in the original  geometry $R^{1,d-1}$ $i.e.$ $\rho_{\rm therm}=U \rho_A U^{-1}$. More explicitly, we may write the density matrix on $D(A)$ as 
\begin{align}
\rho_A=U^{-1}\frac{e^{-H/T_0}}{Z(T_0)}U\,,\qquad \qquad Z(T_0)=\tr(e^{-H/T_0})\,.
\end{align}
By considering the $n$ copies of $\rho_A$ and after taking its trace
\begin{align}
tr[\rho^n_A]=\frac{e^{-nH/T_0}}{Z(T_0)^n}=\frac{Z(T_0/n)}{Z(T_0)^n}.
\end{align}
Now by using the definition of free energy, $F(T)=-T\log(Z(T))$, we can write the R\'enyi entropy as follow
\begin{align}
S_n=\frac{1}{1-n}\log\left(\frac{Z(T_0/n)}{Z(T_0)^n}\right)=\frac{n}{(1-n) T_0}\left\{F(T_0)-F(\frac{T_0}{n})\right\}.
\end{align}
Finally by utilizing the thermodynamical identity $S=-\partial F / \partial T$, one can rewrite the R\'enyi entropy of the above equation to
\begin{align}
S_n=\frac{n}{(n-1) T_0}\int^{T_0}_{T_0/n}S_{\rm therm}(T)dT\,.
\label{renyyi}
\end{align}
Here, $S_n$ is the desired entanglement R\'enyi entropy between the two subsystems $A$ and its complement in the vacuum of the CFT, and $S_{\rm therm}(T)$ denotes the thermal entropy of the CFT.
Upon taking the $n\to 1$ limit of the above formula, $S_{EE}=\lim_{n\to 1}S_n=S_{\rm therm}(T_0)$, which means that the conformal transformations and the corresponding unitarity transformation of the density matrix discussed  above, relate the EE of the SES to the thermal entropy of the same CFT at a temperature $T_0$ in $R^1\times H^{d-1}$ space.

This insight might not be particularly useful in the computation of the EE for a generic CFT unless using the AdS/CFT correspondence. We can relate the thermal bath in the boundary to a topological black hole in the AdS bulk space whose event horizon has a hyperbolic cross-section \cite{Aminneborg:1996iz,Brill:1997mf,Vanzo:1997gw,Mann:1996gj,Birmingham:1998nr,Emparan:1998he}.
According to the AdS/CFT dictionary 
\begin{align}
S_{\rm therm}(T)\Big|_{{\rm CFT\; in\;} R^1 \times H^{d-1} {\;\rm{geometry}}}=
S_{\rm therm}(T)\Big|_{\rm hyperbolic\; BH}\,.
\end{align}
The right hand side of this equation is easy to compute by using the Wald's formula for entropy \cite{Wald:1993nt,Iyer:1994ys} in any general gravitational theory.

The authors of \cite{Hung:2011nu}, use this method to find the R\'enyi entropies of the dual CFTs by finding the thermal entropy of black holes in the Einstein, Gauss-Bonnet (GB)  and quasi-topological gravitational theories.
Moreover, the authors of \cite{Caputa:2013eka, Belin:2013uta} expand the aforementioned method to compute the ERE for the grand canonical ensembles. For these cases, one requires to consider a $U(1)$ charged hyperbolic black hole. 
The computation of the charged R\'enyi entropies in \cite{Pastras:2014oka, Pastras:2015mza} shows that the R\'enyi entropy inequalities put a new restriction on the allowed values of the coupling constants of the GB gravitational theory.
The R\'enyi entropies  obey the following inequalities \cite{karol}
\begin{subequations}
	\begin{eqnarray}
	\frac{\partial S_n}{\partial n} &\leq &0 \,,\label{1NEQ} \\ 
	\frac{\partial}{\partial n} \big(\frac{n-1}{n} S_n \big) & \geq & 0 \,, \label{2NEQ} \\
	\frac{\partial}{\partial n} \big( (n-1)S_n \big) &\geq &0\,, \label{3NEQ} \\ 
	\frac{\partial^2}{\partial n^2} \big( (n-1)S_n \big) & \leq & 0 \,. \label{4NEQ}
	\end{eqnarray}
\end{subequations}
As discussed in \cite{Hung:2011nu}, while the second and the third inequalities  are yielded from the positivity of the black hole thermal entropy  the first and the last inequalities are correct as long as the black hole has a positive specific heat. For a recent study of these inequalities see \cite{Nakaguchi:2016zqi}.

In \cite{Belin:2013dva} it was shown that adding a  scalar field to the Einstein gravitational action creates a ``hairy" black hole solution below a critical temperature and this encourages one to study the R\'enyi entropies of the dual CFT at this phase transition. It was also shown that at this critical temperature where the scalar field condenses, the second derivative of the R\'enyi entropy becomes discontinuous.
Similar phase transitions occur in the study of the holographic superconductors \cite{Hartnoll:2008kx}, where it might cause a phase transition in the dual boundary CFT. For a review on holographic superconductors see \cite{Horowitz:2010gk}.

Inspired by the works of \cite{Belin:2013dva} and \cite{Pastras:2015mza} we are going to add a scalar field to the GB gravity to study the phase transition in the presence of the new gravitational coupling. We also find new restrictions on the allowed region of the GB coupling constant by studying the R\'enyi entropy inequalities.

The organization of the paper is as follow: In section 2 we introduce the Einstein and the hairy black holes and compute their physical quantities such as the temperature, energy, and thermal entropy and also discuss the condensation of the scalar field. We review the related holographic R\'enyi entropy and its behavior under the phase transition. In section 3 we add the Gauss-Bonnet terms and investigate the effect of these higher derivative terms to the results of section 2. In the last section, we discuss our results and the inequalities of the HRE.


\section{Phase transition and R\'enyi entropy}
In this section we review the phase transition between Einstein and hairy black holes in five dimensions which have been already studied in \cite{Belin:2013dva} and \cite{Dias:2010ma}. We also look at the behavior of the R\'enyi entropies under the scalar field condensation.
\subsection{The Einstein black hole}
We begin this section by introducing the Einstein-Hilbert action together with a cosmological constant term in a five  dimensional space-time
\begin{align}
I= \frac{1}{16 \pi G_N} \int d^5 x \sqrt{-g} \Big( \mathcal{R}+ \frac{12}{L^2} \Big)\,.
\label{act1}
\end{align}
To study the entanglement entropy of a spherical region with radius $R$ in a quantum field theory side, which is supposed to exist as a gauge theory on $R^1\times H^{d-1}$ geometry, we need to consider a black hole in the AdS space-time with characteristic scale of $L$ and a hyperbolic spatial boundary. Therefore we start from the following metric 
\bea
ds^2= -N(r)^2 F(r) dt^2 +\frac{dr^2}{F(r)} +r^2 d\Sigma_3^2\,,
\label{eq.2.1.2}
\eea
where
$d\Sigma_3^2$
is the metric on a unit hyperboloid in three dimensions
\begin{align}
d\Sigma_3^2 =du^2 +\sinh^2u \, (d\theta^2 +\sin^2 \theta d\phi^2)\,.
\end{align}
By inserting (\ref{eq.2.1.2}) into the equations of motion, we find the following solutions for the unknown functions of the metric \cite{Dias:2010ma}
\begin{align}
F(r)=\frac{r^2}{L^2} f(r) -1\,,\qquad f(r)=1+\frac{r_H^2 L^2 -r_H^4}{r^4}\,, \qquad N(r)=N=\frac{L}{R}\,,
\label{eq.2.1.6}
\end{align}
where $r_H$ is the radius of horizon.
In a series of straightforward steps, we will compute the temperature, energy and thermal entropy of this black hole. 
 By knowing the metric we may find the temperature of the black hole by using the following definition
\begin{align}
T= \frac{N}{4\pi} \Big[ \frac{d}{d r} F(r) \Big]_{r=r_H}
= \frac{1}{2\pi R}(2 r_H L^{-1} -L r_H^{-1} )\,.
\label{eq.2.1.9}
\end{align}
To compute the black hole energy (mass) in an asymptotically AdS space-time, one could use the Astekhar-Das formalism \cite{Ashtekar:1999jx}. A quick but equivalent way is to expand the $g_{00}$ in the metric \cite{Dias:2010ma}
\begin{align}
g_{00} (r) = -N^2  \left(\frac{r^2}{L^2} -1 -\frac{m}{r^2} +\cdots \right)\,.
\label{eq.2.2.11}
\end{align}
For simplicity we assume that $R=L$ (we can get rid of this coefficient by a rescaling of time).
Therefore the black-hole energy would be
\begin{align}
E= \frac{3 V_{\Sigma}}{16 \pi G_N}m=
\frac{3 V_\Sigma}{16 \pi G_N} \left( r_H^4 L^{-2} -r_H^2 \right)\,,
\label{eq.2.1.13}
\end{align}
where $V_{\Sigma}$ is the regularized volume of the hyperbolic space.
Finally to compute the thermal entropy we use the Wald's formula for entropy
\cite{Wald:1993nt,Iyer:1994ys}
\begin{align}
S_{\text{therm}} = -2\pi \int_{\text{horizon}}\!\!\!\! d^3 x \sqrt{h} \frac{\partial \mathcal{L}}{\partial R_{abcd}} \hat\varepsilon_{ab} \hat\varepsilon_{cd}\,.
\label{eq.2.1.15}
\end{align}
In this formula $h$ is the determinant of the induced metric on the horizon and
$\hat\varepsilon_{ab}$
is a binormal, composed from the Killing vectors $\xi_b$ and the normal vectors to the horizon $\eta_a$, $i.e.$
$\hat\varepsilon_{ab} =\eta_a \xi_b -\eta_b \xi_a$.
At the level of the Einstein-Hilbert action, the Wald's formula coincides with the Bekenestein-Hawking area formula
\begin{align}
S_{\text{therm}}=\frac{\mathcal{A}_{\text{horizon}}}{4G_N}= \frac{r_H^3}{4}  \frac{V_\Sigma}{G_N} 
\label{eq.2.1.18}\,.
\end{align}
For future purposes it would be useful to introduce  a dimensionless parameter $x\equiv\frac{r_H}{L}$, so that the temperature, energy and thermal entropy of the black hole can be written as
\begin{align}
\tilde T=\frac{1}{2\pi} \big(2x -\frac{1}{x} \big)\,, \qquad
\tilde E= \frac{3}{16 \pi} \, x^2 (x^2-1)\,,\qquad
\tilde S=\frac{x^3}{4}\,,
\label{TESx}
\end{align}
where we have assumed $R=L=\frac{V_\Sigma L^2}{G_N}=1$ for simplicity. It is easy to show that the first law of thermodynamics, $i.e.$ $d\tilde{E}=\tilde{T} d\tilde{S}$, is satisfied by the above values.

\begin{figure}[!htbp]
\centering
\includegraphics[width=0.5\textwidth]{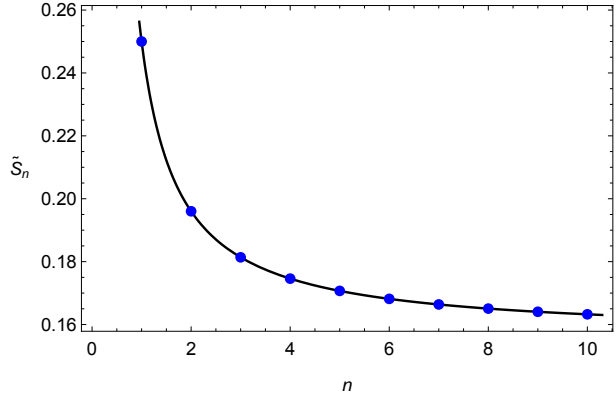}
\caption{The R\'enyi entropies computed from the hyperbolic black hole solution}
\label{fig1}
\end{figure}

Now we can go one step further and find the R\'enyi entropies  which we have introduced in equation (\ref{renyyi}) by integrating the thermal entropy
\begin{align}
S_n= \frac{n}{n-1} \int_{\frac{T_0}{n}}^{T_0} S_\text{therm} dT=
\frac{n}{T_0 (n-1)} \int_{x_n}^1 \tilde{S}(x) \frac{d\tilde{T}}{dx} dx=\frac18\frac{n}{n-1} \left( 2-x_n^2 (x_n^2 +1) \right)\,,
\label{eq.2.1.20}
\end{align}
where $T(x) \big|_{x=x_n} =\frac{T_0}{n}$ and $T_0$ is related to the length scale of the hyperbolic space, here $T_0=\frac{1}{2\pi}$.
In order to draw the behavior of the $S_n$ versus $n$, we should substitute the value of $x_n$ in $S_n$. It is given by the real positive root of the following quadratic equation
\begin{align}
\frac{1}{2\pi} ( 2x_n -\frac{1}{x_n} ) =\frac{1}{2\pi n}
\Rightarrow x_n =\frac{1}{4n} ( 1+\sqrt{1+8n^2} )\,.
\label{eq.2.1.22}
\end{align}
Substitution gives
\be
\tilde S_n=\frac{40 n^4-12n^2-1 - (1+8n^2)\sqrt{1 + 8 n^2}}{
256 (n-1) n^3}\,.\label{snebh}
\ee
In the limit $n\rightarrow 1$, $\tilde{S}_1=\frac14$ gives the entanglement entropy and when $n\rightarrow\infty$ the R\'enyi entropy goes to the $\tilde{S}_\infty\rightarrow \frac{5}{32}$ (see figure \ref{fig1}).
 Moreover one can simply check that all the inequalities in (\ref{1NEQ})-(\ref{4NEQ}) hold for the R\'enyi entropy in (\ref{snebh}).

\subsection{The hairy black hole}
In this section we review the scalar field condensation in the background of an AdS black hole with a hyperbolic spatial slicing. We also study the effect of condensation (phase transition) on the behavior of the R\'enyi entropies. To begin, we add a real scalar field $\psi$ which is coupled to gravity as
\begin{align}
I= \frac{1}{16 \pi G_N} \int d^5 x \sqrt{-g} \left( \mathcal{R}+ \frac{12}{L^2} -\mu^2 \psi^2 - (\nabla \psi)^2 \right)\,.
\label{act2}
\end{align}
For simplicity we just consider a mass and a kinetic term.
To find the desired black hole solution, we choose the following ansatz for the metric and scalar field
\begin{align}
ds^2= -F(r) e^{2\chi (r)} dt^2 +\frac{dr^2}{F(r)} +r^2 d\Sigma_3^2\,, \qquad \psi=\psi(r)\,.
\label{eq.2.2.2}
\end{align}
The back reaction of the scalar filed on the metric will appear through $F$ and $\chi$ functions. The equations of motion are
\begin{align}
&\psi'' (r) +\frac{1}{3L^2 r F(r)} \big( \psi' (r) [ 6 (2r^2-L^2) +3 L^2 F(r) -r^2 \mu^2 L^2 \psi (r)^2 ] -3r\mu^2 L^2 \psi(r) \big)=0\,,\notag\\
& F'(r) +F(r) \big( \frac{2}{r} +\frac{r}{3} \psi'(r)^2 \big) +\frac{r}{3} \mu^2 \psi(r)^2 +\frac{2}{r} -\frac{4r}{L^2} =0\,,\notag\\
& \chi'(r) +\frac{r}{3} \psi'(r) ^2 =0\,,
\label{eom1}
\end{align}
where the prime stands for differentiation with respect to the radial coordinate $r$.
To solve this system of  coupled nonlinear differential equations numerically, we need to impose boundary conditions on the metric and scalar field at the horizon ($r=r_H$) and at the space-like  boundary ($r \to \infty$) as follows:

$\bullet$ $F(r):$
 The definition for the location  of the horizon demands that
$F(r_H) =0$.
For asymptotically AdS black holes and at large distances near the infinite boundary, $F(r)$ must behave as
$\frac{r^2}{L^2}-1$.

$\bullet$ $\psi (r):$ 
In order to have a hairy black hole, we need a regular scalar field at the horizon of the black hole, so we suppose that
$\psi (r_H) =\mathcal{O}(1)$. As is well known in the dictionary of AdS/CFT, our gravitational bulk theory must be dual to a conformal field theory that lives at the boundary of the AdS space. This CFT contains a scalar operator with the conformal dimension $\Delta$ so that $\mu^2 L^2=\Delta (\Delta -4)$. In order to have an asymptotically AdS black hole, unstable against the scalar field condensation, $\Delta$ has to be sufficiently small. We use the benefits of this instability to study the phase transition due to the scalar field condensation.  The instability condition requires that the scalar mass takes the values between the two Breitenlohner-Freedman (BF) bounds of $\text{AdS}_5$ and $\text{AdS}_2$
\cite{Breitenlohner:1982jf, Mezincescu:1984ev, Belin:2013dva}
\begin{align}
-4 \le \mu^2 L^2 \le -1\,.
\label{eq.2.2.4}
\end{align}
To specify the behavior of the scalar field at the AdS boundary, it will be sufficient to study its equation of motion around the AdS-Schwarzschild background, which assumes the form of a Klein-Gordon equation. The solution to this equation behaves asymptotically as
\begin{align}
\psi(r) \approx \frac{A_{(+)}}{r^{\Delta_+}} +\frac{A_{(-)}}{r^{\Delta_-}}\,,
\label{eq.2.2.5}
\end{align}
where $\Delta_{\pm}=2 \pm \sqrt{4+ \mu^2 L^2}$, and $A_{(\pm)}$ are the expectation values of the conformal operators with conformal dimensions $\Delta_\pm$ , $i.e.$ $\langle O_{\Delta_{\pm}} \rangle \equiv A_{(\pm)} $.
If we assume the Dirichlet boundary condition then we can keep the fastest falling off mode near the boundary and ignore the other one, hence we choose, $A_{(-)}=0$.

$\bullet$ $\chi (r):$
By using the third equation of (\ref{eom1}) and applying the boundary conditions $\psi(r) \approx \frac{A_{(+)}}{r^{\Delta_+}}$ and $\psi(r_H)=\mathcal{O}(1)$, it is easy to obtain the boundary conditions for
$\chi (r)$
\begin{align}
\chi (r\to \infty) = \mathcal{O} (r^{-2 \Delta_+} ) +\cdots\,, \qquad \chi (r_H) =\mathcal{O} (1)\,.
\label{eq.2.2.7}
\end{align}

In this paper, we are going to find the numerical solutions for equations of motion by using the shooting method. To do this it would be easier to change the radial variable from $r$ to $z$ via $z=\frac{r_H}{r}$. This substitution maps the region $r_H \le r < \infty$  to the region
$1\ge z>0$.
Here we also introduce two dimensionless parameters,
$m=-\mu^2 L^2$ and $z_0=\frac{r_H}{L}$. By these changes the equations of motion assume the form
\begin{align}
&\psi'' (z) + \frac{1}{z} \psi' (z) \big(1 +\frac{1}{F(z)}
 [2 -\frac{z_0^2}{z^2}(4+\frac{m}{3} \psi(z)^2) ] \big) +m\frac{\psi (z)}{F(z)} \frac{z_0^2}{z^4} =0\,, \notag \\
&F'(z) -F(z) \big( \frac{2}{z} +\frac{z}{3} \psi'(z)^2 \big) +\frac{z_0^2}{3 z^3} m\psi(z)^2 -\frac{2}{z} +\frac{4z_0^2}{z^3} =0\,, \notag \\
&\chi' (z) +\frac{z}{3} \psi'(z)^2=0\,,
\label{eom2}
\end{align}
and boundary conditions yield
\begin{alignat}{2}
&F_{B} (z \to 0)  \sim \frac{z_0^2}{z^2} -1\,,&&\qquad F_{H} (z \to 1) =0\,, \notag \\
&\psi_{B} (z \to 0) \sim C_+ z^{\Delta_+} \,,
&&\qquad \psi_{H} (z \to 1) = \mathcal{O} (1)\,, \notag \\
&\chi_{B} (z \to 0) \sim \mathcal{O} (z^{2\Delta_+} )\,, &&\qquad \chi_{H} (z \to 1) =\mathcal{O} (1)\,.
\label{BDCz}
\end{alignat}
In the shooting method, we need to calculate the series expansions of the scalar and metric functions around the horizon and boundary. By inserting these series into the equations of motion we can find the unknown coefficients of the expansions. After that, we match the series  smoothly at some small distance, say $\epsilon$, from one end of the interval $0 < z <1$. Note that for a given $\psi (z)$ one can find $\chi (z)$
from the last equation of motion in  (\ref{eom1}), simply by an integration. 

Since the equations of motion are regular at the horizon, the
expansion of functions are given by a Taylor series near the horizon 
\begin{equation}
\begin{aligned}
\psi_{H} (z)=\sum_{n=0}^\infty \psi_{(n)} (z-1)^n\,, \qquad
F_{H} (z) =\sum_{n=0}^\infty F_{(n)}  (z-1)^n\,.
\end{aligned}
\label{eq.2.2.12}
\end{equation}
The boundary conditions \eqref{BDCz} require that the $F_{(0)}=0$. By substituting the near horizon expansions of \eqref{eq.2.2.12} into the equations  of motion \eqref{eom2} and expanding again near the $z=1$, we can find the values of $F_{(n)}$ and $\psi_{(n)}$ as functions of free parameters $z_0$, $m$ and $\psi_{(0)}\equiv \psi_1 $.
In the precision region of our computations it would be enough to keep expansions up to $n=6$. For example the first few terms are\footnote{ It is important to note that by going to the higher orders of expansion bigger than six, the change in our numerical results (graphs) was much smaller than the precision that we have used in our calculations.}
\begin{align}
\psi_{H} (z) & = \psi_1 +(z-1) \big( \frac{3m z_0^2 \psi_1}{z_0^2 (12+m \psi_1^2) -6} \big)+\cdots\,, \notag
\\
F_{H}(z) &=(z-1) \big( 2-4z_0^2 -\frac{m}{3} z_0^2 \psi_1^2 \big) +\cdots\,.
 \label{NHE}
\end{align}
The expansion near the boundary at  $z=0$ is a little bit trickier, because the equations of motion are irregular at this point. Here we have a power series expansion and the power of the leading term depends on the value of $m$. We denote this, by a mass dependent parameter $\delta$
\begin{align}
\psi_{B} (z)=C_{+} z^{\Delta_{+}}+\cdots= \sum_{i=i_{min}}^{\infty} a_i z^{2+i\delta}\,,\qquad 
F_{B} (z)= \frac{z_0^2}{z^2} -1 +\sum_{j=1}^{\infty} b_j z^{j\delta} \,. 
\label{NBE}
\end{align}
Here we have expanded the functions until the resulting algebraic system of equations gives a non-trivial solution.  In table \ref{tab1} we have presented the values of $m, \Delta_{+}, \delta$ and $i_{min}$ which we have used to solve the equations of motion numerically
\begin{table}[!h]
\centering
\setlength{\tabcolsep}{0.2cm}
\renewcommand{\arraystretch}{1.2}
\begin{tabular}{|c||c|c|c|c|c|c|}
\hline
$m$ & $4$ & $\frac{63}{16}$ & $\frac{60}{16}$ & $\frac{55}{16}$ & $3$ & $\frac{39}{16}$ \\ \hline
$\Delta_+$ & $2$ & $\frac{9}{4}$ & $\frac{10}{4}$ & $\frac{11}{4}$ & $3$ & $\frac{13}{4}$  
\\ \hline
$\delta$ & $1$ & $\frac{1}{4}$ & $\frac{1}{2}$ & $\frac{1}{4}$ & $1$ & $\frac{1}{4}$ \\\hline
$i_{min}$ & $0$ & $-1$ & $-1$ & $-3$ & $-1$ & $-5$ \\\hline
\end{tabular}\caption{The numerical values for series expansion near the boundary.}\label{tab1}
\end{table}

By substituting the expansions in (\ref{NBE}) for every mass parameter $m$ from table 1 into the equations of motion, we can find the expansion coefficients,  $a_i$ and $b_j$ in terms of three free parameters $C_{+}, C_m$ and $z_0$. Here $C_{+}$ is the coefficient of $z^{\Delta_+}$ in $\psi_{B} (z)$ and $C_m$ is the coefficient of $z^2$ in $F_{B} (z)$ in equation (\ref{NBE}) (note that $C_m z^2$ always exists in the near boundary expansion of $F(z)$). 

Let us explain the shooting method a little bit. At first step, we read the initial values of the fields and their derivatives near the horizon at $z=1-\e$ from the series expansion in (\ref{NHE}), remembering that the coefficients are functions of $C_{+}, C_m$ and $z_0$. Then we solve the first two differential equations in (\ref{eom2})  numerically to find the values of the fields near the boundary at $z=\e$ (note that we need the very small parameter $\e$ as a regulator).

Now we can compare these values of the fields (two equations) and their first derivatives (two equations) with those which are computed from the near boundary expansion in (\ref{NBE}). At the end of the day, we will have four equations for four unknown parameters $C_{+}, C_m, \psi_1$ and $z_0$ to solve. 
This set of equations has innumerable numerical solutions which can be found and put into the power series expansions of the fields both at the horizon (\ref{NHE}) and boundary  (\ref{NBE}) to compute various physical and thermodynamical quantities. 

The first quantity that we consider, is again the temperature of the hairy black hole 
\begin{align}
T =\frac{e^{\chi (r)}}{4\pi} |F'(r)|\Big|_{r=r_H}=\frac{e^{\chi (z)}}{4\pi z_0} |F'(z)|\Big|_{z=1}\,,
\label{eq.2.2.18}
\end{align}
where $F'(1) =2-4 z_0^2 -\frac{m}{3} z_0^2 \psi_1^2$ is computed from equation \eqref{NHE} and
\begin{align}
\chi(1) = \int_{z=\epsilon}^{1-\epsilon} (-\frac{z}{3} ) (\psi'(z) )^2 dz \,,
\label{eq.2.2.20}
\end{align}
can be computed numerically by imposing the boundary conditions of (\ref{BDCz}).

The next quantity is the thermal entropy of the hairy black hole 
\begin{align}
S_{\text{therm}} =\frac{r_H^3}{4} \frac{V_\Sigma}{G_N} \rightarrow \tilde{S}=\frac{z_0^3}{4}\,,
\label{eq.2.2.21}
\end{align}
where  we have calculated the entropy again in units where $\frac{L^3 V_\Sigma}{G_N}=1$. 

To compute the energy of the hairy black hole we can use two different ways again. Either we can expand the $g_{00}$ 
\begin{align}
g_{00} (z)_{z \to 0}= -F(z)e^{2\chi(z)}\Big|_{z\rightarrow 0}=  -\frac{z_0^2}{z^2} +1 +\tilde m z^2 +\cdots\,,
\label{eq.2.28}
\end{align}
and pick up the coefficient of $\frac{z^2}{z_0^2}$, therefore the energy yields
\begin{align}
E=\frac{3 V_\Sigma}{16 \pi G_N} \tilde m z_0^2 \rightarrow \tilde{E}=\frac{3}{16\pi} \tilde m z_0^2\,,
\label{eq.2.29}
\end{align}
or we can use the first law of thermodynamics for black holes ($dE=T dS$) numerically and compute the energy. Both ways provide the same result.

As we discussed previously, in the AdS/CFT dictionary the coefficient of $r^{-\Delta_{+}}$ gives the expectation value of the conformal operator $O$ with a conformal dimension $\Delta_{+}$. This coefficient is called the condensate.
In the $z$ coordinates at leading order of $\psi(z) \sim C_{+} z^{\Delta_+} $ therefore the value of the condensate is $\langle O \rangle =C_{+} z_0^{\Delta_+}$. 

\begin{figure}[!htbp]
\centering
\includegraphics[width=1\textwidth]{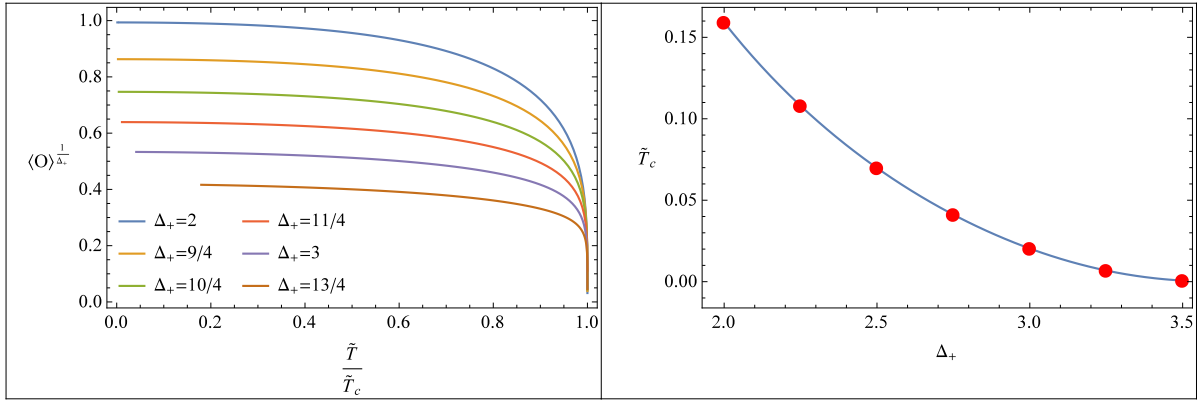}
\caption{\footnotesize{Left: The condensate as a function of temperature (scaled by the corresponding critical temperature) for various values of $\Delta_+$. The lowest curve corresponds to $\Delta_+=\frac{13}{4}$ and the top curve is for $\Delta_+=2$. Right: The critical temperature as a function of $\Delta_+$.}}\label{fig2}
\end{figure}

Figure \ref{fig2} shows the behavior of the $\langle O \rangle^\frac{1}{\Delta_{+}}$ in terms of the temperature, for various  possible values of the $\Delta_{+}$ in the interval  $2 \leq \Delta_{+} \leq 2+\sqrt{3}$. The condensation happens for each value of the $\Delta_{+}$ at different critical temperatures $\tilde T_c$, which decreases when $\Delta_{+}$ increases.

When the $\text{AdS}_5$ BF bound saturates, $i.e.$ $\Delta_+ =2$ or $ m =4$, the critical temperature $T_c$ reaches to its maximum value at $\tilde T_0 =\frac{1}{2\pi}\approx 0.159$. This is the Rindler temperature (the temperature of massless black hole).

In the following figures, we have sketched the behavior of various physical quantities as a function of the energy or the temperature.

\begin{figure}[!htbp]
\centering
\includegraphics[width=1\textwidth]{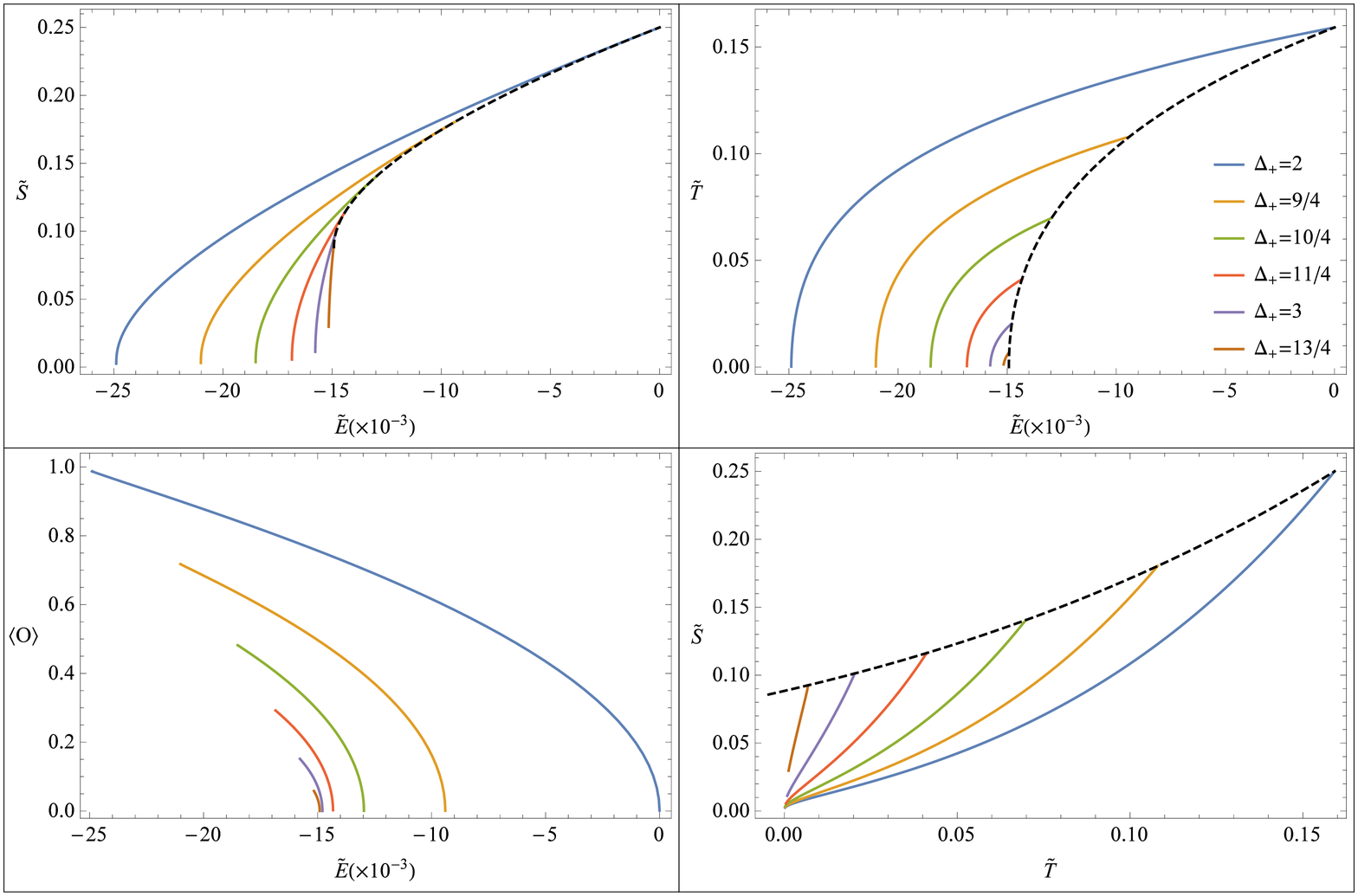}
\caption{\footnotesize{Thermodynamical quantities for various values of $\Delta_+$. Note that the colors in this figure match with those of figure \ref{fig2}. The dashed black curve represents the Einstein black hole solution and the solid curves belong to the hairy black holes.}}
\label{fig3}
\end{figure}

In each diagram in figure \ref{fig3} the dashed curve represents the Einstein black hole. The $\tilde{S}-\tilde{T}$ diagram shows how the condensation happens. When temperature decreases there is a critical temperature for each value of the $\Delta_{+}$ (the point where solid curves meet the dashed one) in which, the Einstein black hole is replaced by a hairy black hole. As we see the critical temperature decreases when $\Delta_{+}$ increases. One can convert the temperature to energy using the $\tilde{T}-\tilde{E}$ diagram. 
The  $\tilde{S}-\tilde{E}$ diagram shows that the entropy of a hairy black hole at a specific value of the energy is larger than that of the Einstein black hole, therefore, the hairy black holes are more favorable (stable) after the condensation. The $\langle O\rangle-\tilde{E}$ diagram displays where the condensation starts and reaches to its maximum value in terms of the energy.

\begin{figure}[!htbp]
\centering
\includegraphics[width=1\textwidth]{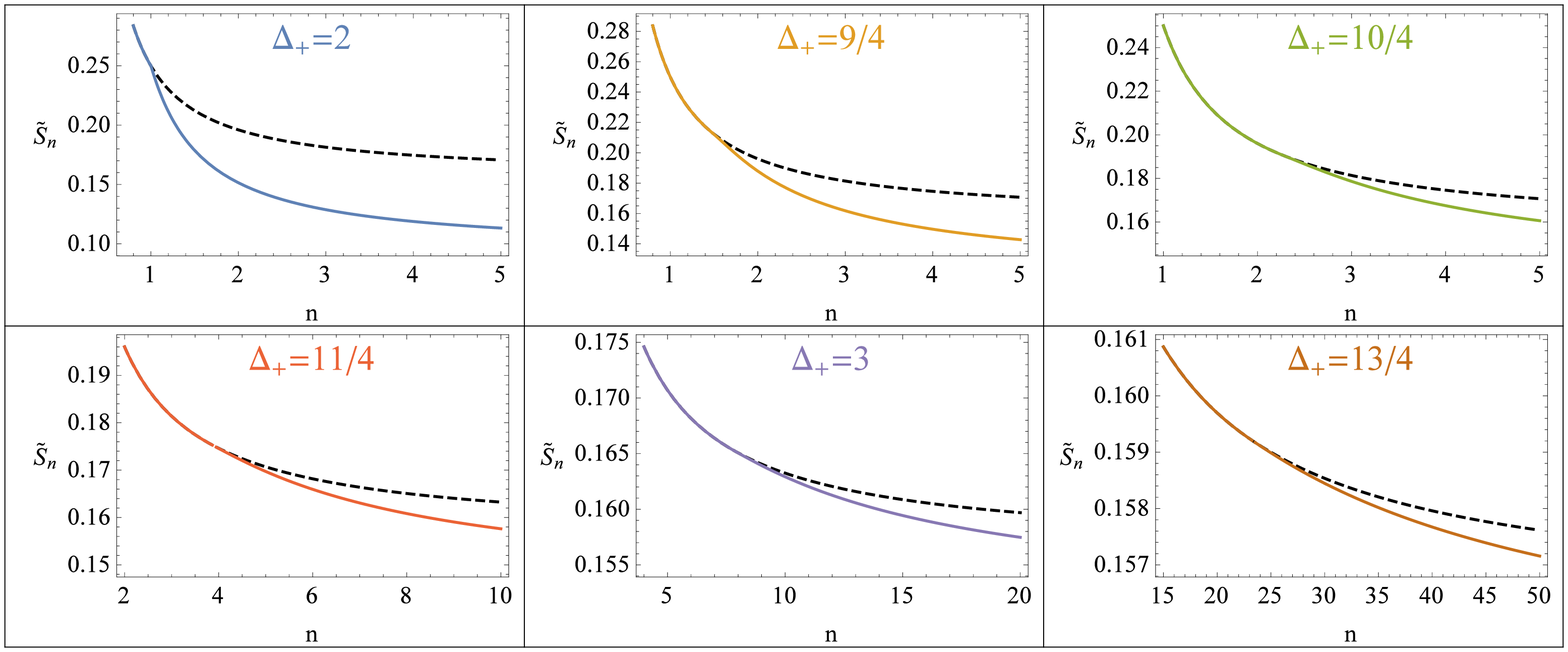}
\caption{\footnotesize{Each diagram shows the behavior of the R\'enyi entropies for EBH (dashed curve) and HBH for various values of the $\Delta_{+}$. Two curves meet each other at a critical number, $n_c$.}}
\label{fig4}
\end{figure}

To compute the R\'enyi entropies from (\ref{renyyi}) according to the $\tilde{S}-\tilde{T}$ diagram in the figure \ref{fig3}, we should take into account the phase transition from the Einstein black hole (EBH) to the hairy black hole (HBH) \cite{Belin:2013dva}
\begin{align}
S_n =\frac{n}{T_0 (n-1)} \left\{ \int_{T_0/n}^{T_c} S_{\text{therm}}^{\text{HBH}} (T) \,dT +\int_{T_c}^{T_0} S_{\text{therm}}^{\text{EBH}} (T)\, dT \right\}\,.
\label{eq.2.2.28}
\end{align}
The second term above, differs from \eqref{eq.2.1.20} only on its lower limit. Here we can define $T(x) \big|_{x=x_c}=T_c$ in accordance with the definition of the $x_n$. So the $\tilde S_n$ yields
\begin{align}
\tilde S_n &= \frac{n}{\tilde T_0 (n-1)} \left\{ \int_{\tilde T_0/n}^{\tilde T_c} S_{\text{therm}}^{\text{HBH}} (T) \, dT +\int_{x_c}^{1} \tilde S^{\text{EBH}} (x) \frac{d \tilde T}{dx} dx \right\} \notag \\
&= \frac{2\pi n}{n-1} \int_{\tilde T_0/n}^{\tilde T_c} S_{\text{therm}}^{\text{HBH}} (T)\, dT + \frac{n}{8(n-1)} [ 2-x_c^2 (x_c^2 +1)] \,,
\label{eq.2.2.29}
\end{align}
where $x_c$ is the real positive root of the $\frac{1}{2\pi} \left( 2 x_c -\frac{1}{x_c} \right) =\tilde T_c $ or $ x_c =\frac{\pi \tilde T_c}{2} \Big( 1+ \sqrt{1+\frac{2}{(\pi \tilde T_c)^2}}\Big)$.
By a numerical computation, we can draw the $\tilde S_n$ in terms of $n$, see figure \ref{fig4}.

\begin{figure}[!htbp]
\centering
\includegraphics[width=0.6\textwidth]{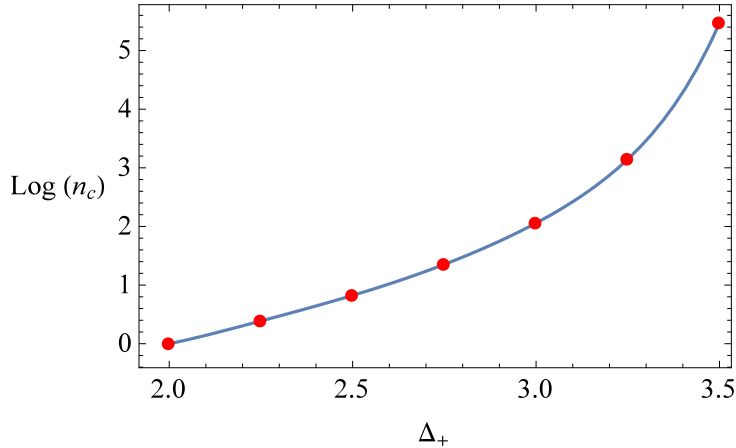}
\caption{\footnotesize{The logarithm of the critical R\'enyi parameter as a function of $\Delta_+$.}}
\label{fig5}
\end{figure}

As we observe in figure \ref{fig4}, one can define a critical number $n_{\text{c}} =\frac{\tilde T}{\tilde T_c}$, where the curve of HBH approaches that of EBH. At this point the second derivative of $\tilde S_n$ suddenly changes. This discontinuity of the second derivative of $\tilde S_n$ with respect to $n$ confirms that the condensation of the scalar field is a second order phase transition. Moreover, the value of $n_c$ increases when $\Delta_{+}$ increases as we depicted in figure \ref{fig5}.
Here all four inequalities (\ref{1NEQ})-(\ref{4NEQ}) hold again.


\subsection{A note on the normalizability of the scalar modes}
The scalar modes that we have considered so far, are constant on the hyperboloid $i.e.$ $\psi=\psi (r)$. These modes are called the non-normalizable modes in the context of AdS/CFT. Since these modes preserve the symmetries of the hyperboloid, the construction of fully backreacting hairy black holes in the gravitational background is relatively easy; and for this reason, they offer a good toy model in the study of the phase transition via the condensation of a scalar field in the gravitational theory.

However, the more physical cases would be those which are not constant on the hyperboloid $\psi=\psi (t,r,u,\theta,\phi)$, $i.e.$ the normalizable modes. In these cases, the construction of the fully backreacting hairy black holes is considerably more difficult (if not impossible). As an example, the authors of \cite{Belin:2013dva} perform a linearized analysis of the Klein-Gordon equation for these modes and manage to show the existence of a phase transition. The main result of this linearized analysis is a phase transition, that occurs if the scalar mass was in the range $-4 \le \mu^2 L^2 \le -3$ which is more restricting than $-4 \le \mu^2 L^2 \le -1$. 

Conversely, one could say that the linearized analysis restricts the admissible region of conformal dimension from $2 \leq \Delta \leq 2+\sqrt{3}$ to $1 \leq \Delta \leq 3$. Indeed, if one compares the figure \ref{fig5} of our paper to the figure 6 in \cite{Belin:2013dva}, one would see that both curves generally behave similarly, especially in the lower admissible region of the conformal dimension. The reason for the similarity of these two curves is hidden in the fact that if the scalar mass is just above the $\text{AdS}_{5}$ BF bound and  below the unitarity bound (a bound imposed by unitarity conditions) $i.e.$ $-\frac{4}{L^2}<\mu^2 <-\frac{3}{L^2}$, then the constant modes on the hyperboloid are normalizable, because it is possible to quotient the hyperboloid to form a compact space, see for example \cite{Aros:2002te, Klebanov:1999tb, Belin:2013dva, Dias:2010ma}. 

Therefore, despite the difference in the length of the allowed region and the difference in the behavior of the aforementioned curves in the upper part of the allowed region, at least in the lower parts, the constant modes of the scalar field could be regarded as a good approximation to the complicated problem of finding the fully backreacting normalizable modes.

All in all, since our main objective in this paper is to study the effect of the higher derivative terms on the entanglement R\'enyi entropy in the process of the phase transition, we will consider the constant modes as a good toy model and focus on them in the remainder of this paper. We will leave the study of normalizable modes to future works.


\section{The ERE from the Gauss-Bonnet gravity} 
In this section, we are going to explore the effect of higher curvature bulk theories of gravity on the condensation and the holographic R\'enyi entropy. We consider the Gauss-Bonnet (GB) gravity in five dimensions where the equations of motion are of the second order in derivatives and the numerical method of the previous section works here as well without imposing any further boundary condition. The GB gravity has been studied extensively in the literature \cite{Cai:2001dz, Cvetic:2001bk, Anninos:2008sj, Bhattacharjee:2015qaa, Ge:2008ni}, which we will review in the following subsection. Afterwards, in subsection 3.2, we will present our own original work on the study of the hairy black holes for this gravitational theory.
\subsection{The modified Einstein Gauss-Bonnet black hole}
Similarly to the previous section, we start from a pure gravitational action in five dimensions
\begin{align}
I=\frac{1}{16 \pi G_N} \int d^5 x \sqrt{-g} \Big\{ \mathcal{R}+\frac{12}{L^2} +\frac{\lambda L^2}{2} \Big(\mathcal{R}_{abcd} \mathcal{R}^{abcd} -4\mathcal{R}_{ab} \mathcal{R}^{ab} +\mathcal{R}^2\Big) \Big\}\,, \label{eq.3.1.1}
\end{align}
where $\l$ is a free dimensionless coupling.
To find the asymptotically AdS black hole solutions with a hyperbolic horizon, we use again the ansatz in equation \eqref{eq.2.1.2} and insert it into the equations of motion. We will find
\begin{align}
\Big(1-\frac{2\lambda L^2}{r^2} \big(1+F(r) \big)\Big)F'(r)+ \frac{2}{r}\big(1+F(r)\big) -\frac{4r}{L^2}  =0\,,
\qquad
\frac{dN(r)}{dr}=0\,,
\label{eq.3.1.3}
\end{align}
with the following Einstein Gauss-Bonnet (EGB) black hole solution
\begin{align}
F(r)&= \frac{r^2}{L^2} f(r) -1 =\frac{r^2}{L^2} \Big( \frac{1}{2\lambda} \Big[ 1- \sqrt{1-4\lambda \Big( 1 -\frac{C}{r^4} \Big)} \Big]\Big) -1\,,
\label{eq.3.1.4}
\end{align}
where the constant of integration is fixed in terms of the horizon radius by demanding that $F(r)$ vanishes at the horizon, $i.e.$ $C=r_H^4 -r_H^2 L^2 +\lambda L^4$.
By comparing \eqref{eq.3.1.4} with equation \eqref{eq.2.1.6} we can fix the constant value of $N(r)=N$. Since asymptotically 
$f(r) \rightarrow f_{\infty}=\frac{1}{2\lambda} (1- \sqrt{1-4\lambda})$ we can choose
\begin{align}
N=\frac{L \sqrt{2 \lambda}}{R \sqrt{(1- \sqrt{1-4\lambda}})} \equiv \frac{L_{\text{eff}}}{R}\,,
\end{align}
where $L_{\text{eff}}$ is an effective asymptotic AdS scale.
Note that asymptotically 
$ F(r) \rightarrow \frac{r^2}{L_{\text{eff}}^2} -1$.

In order to avoid an imaginary AdS scale or a naked singularity, the Gauss-Bonnet coupling $\l$ has to be limited to the region $\lambda \le 1/4 $. On the other hand the unitarity of the boundary theory dual to the Gauss-Bonnet gravity in this background, demands that $-\frac{7}{36} \le \lambda \le \frac{9}{100}$, for example see {\cite{Hofman:2008ar, Camanho:2009vw}, or \cite{Buchel:2009sk}.
Therefore we will restrict our numerical computations to this interval from now on.

Now we follow analogous steps, similar to the section two, to calculate the thermodynamical quantities and then compute the holographic R\'enyi entropy. The black hole temperature is given by
\begin{align}
T=\frac{N}{4\pi}  \frac{\partial F(r)}{\partial r} \Big|_{r=r_H}= \frac{L_{\text{eff}}}{4\pi R L^2}  \frac{2 r_H (2 r_H^2 -L^2)}{ r_H^2 -2\lambda L^2} \,.
\end{align}
By introducing the $X =\frac{r_H}{L_{\text{eff}}}$ and $T_0 =\frac{1}{2\pi R}$, the temperature can be expressed as \cite{Hung:2011nu}
\begin{align}
T=\frac{T_0 X (2 X^2 -f_\infty)}{f_\infty (X^2 -2\lambda f_\infty)}\equiv 2\pi T_0\tilde{T}\,.
\label{eq.3.1.8}
\end{align}
To compute the black hole energy, we use again the procedure discussed in the section two. By expanding the metric near the boundary and assuming that $R=L_{\text{eff}}$ we have
\begin{align}
g_{00} (r)\Big|_{r \to \infty} =  -\Big\{ \frac{r^2}{L_{\text{eff}}^2} -1 -\frac{C}{r^2 L^2 \sqrt{1-4\lambda}}+\cdots \Big\}\,.
\end{align}
Therefore the black hole energy assumes the form
\begin{align}
E&=\frac{3V_\Sigma}{ 16 \pi G_N} \Big( \frac{r_H^4 -r_H^2 L^2 +\lambda L^4}{L^2 \sqrt{1-4\lambda}} \Big)\notag \\
&= \frac{3 V_\Sigma L^2}{16 \pi G_N} \frac{ X^4 -f_\infty X^2 +\lambda f_\infty^2}{f_\infty^2 \sqrt{1-4\lambda}} \equiv \Big( \frac{V_\Sigma L^2}{G_N} \Big) \tilde E\,,
\label{eq.3.1.14}
\end{align}
where $\tilde E$ is the dimensionless energy of the EGB black hole. 
As it was mentioned in section two, for the higher derivative gravitational theories we must use the Wald's formula \eqref{eq.2.1.15} to compute the thermal entropy. 
After some algebra we find 
\begin{align}
\frac{\partial \mathcal{L}_{GB}}{\partial R_{abcd}} \hat \varepsilon_{ab} \hat \varepsilon_{cd} =\frac{1}{16 \pi G_N} \Big\{
\frac{12 \lambda L^2}{r^2} [ 1+F(r) ]-2 \Big\},
\label{eq.3.1.20}
\end{align}
from which, we can express the thermal entropy as
\begin{align}
S_{\text{therm}}&
=\frac{V_\Sigma}{4 G_N}  r_H^3 \Big( 1-\frac{6 \lambda L^2}{r_H^3} \Big) \notag \\
&= \frac{V_\Sigma L^3}{G_N}  \frac{X^3 -6 \lambda X f_\infty}{4 f_\infty^{3/2}} \equiv \Big( \frac{V_\Sigma L^3}{G_N}\Big) \tilde S_{\text{therm}}\,.
\label{eq.3.1.21}
\end{align}
By knowing the thermal entropy we can compute the holographic R\'enyi entropy. Using the equation \eqref{eq.2.1.20} and defining the $T(X)\big|_{X=X_n}= \frac{T_0}{n}$, we find \cite{Hung:2011nu}
\begin{align}
S_n&= \frac{n}{T_0 (n-1)} \int_{X_n}^1 S(X) \frac{dT}{dX} =\frac{n}{T_0 (n-1)} \int_{X_n}^1 \Big( \frac{L^3 V_\Sigma}{G_N} \tilde S_{\text{therm}} \Big) \Big( \frac{1}{R}\, \frac{d\tilde T}{dX}\Big) dX
\notag\\&
= \frac{n}{n-1} \frac{L^3 V_\Sigma}{G_N}  \frac{1}{8 f^{5/2}} \Big( 9(1-X_n^4) -3 f_\infty (1-X_n^2) +\frac{4 (f_\infty -2)}{1-2\lambda f_\infty} -\frac{4 X_n^4 (f_\infty -2X_n^2)}{X_n^2- 2\lambda f_\infty} \Big)\,.
\label{eq.3.1.23}
\end{align}
Note that, $X_n$ stands for the real positive root of the following third order equation
\begin{align}
2n X_n^3-f_\infty X_n^2 -n f_\infty X_n +2\lambda f_\infty^2=0\,.
\label{eq.3.1.24}
\end{align}
In \cite{Pastras:2014oka} and \cite{Pastras:2015mza} the authors show that in the Gauss-Bonnet gravity there is a new bound on the coupling $\l$ when we demand the positivity of the thermal entropy. In five dimensions, this is $-\frac{7}{36}\leq \l \leq \frac{1}{12}$. We will discuss this in the next section.

\subsection{The modified hairy black hole}
In section two, we reviewed the condensation of the scalar field around a five-dimensional asymptotically AdS black hole with the hyperbolic horizon in the Einstein gravity. In this section, we are going to study the effect of higher-derivative terms on the scalar condensation, by considering the Gauss-Bonnet gravity.
We first prove the existence of a non-trivial scalar solution below the critical temperature and then compare its thermodynamical properties and  holographic R\'enyi entropy with those of the hairy black hole of the Einstein gravity, in various values of the scalar mass $\mu$ and coupling $\l$.
Let us start from the following action where a real scalar field is coupled to the gravity
\begin{align}
I=\frac{1}{16 \pi G_N}\!\! \int \!\! d^5 x \sqrt{-g} \left\{ \mathcal{R}+\frac{12}{L^2} +\frac{\lambda L^2}{2} \Big(\mathcal{R}_{abcd} \mathcal{R}^{abcd} -4\mathcal{R}_{ab} \mathcal{R}^{ab} +\mathcal{R}^2\Big)\!-\!\mu^2 \psi^2 -(\nabla \psi)^2 \right\}.
\label{eq.3.2.1}
\end{align}
The metric and the scalar field ansatz for the Gauss-Bonnet gravity are the same as those in equation \eqref{eq.2.2.2}. The equations of motion are
\begin{align}
&\psi'' (r) +\frac{1}{3L^2 r F(r) [1-V(\lambda,r)]} \bigg\{ \psi' (r) \big[ 6(2r^2 -L^2) +3L^2 -\mu^2 L^2 r^2 \psi(r)^2 \notag\\&
-9 L^2 F(r)V(\lambda,r) \big] -3\mu^2 L^2 r \psi(r) [ 1-V(\lambda,r)] \bigg\}=0\,,\notag\\
&F'(r)+ \frac{1}{[1-V(\lambda,r)]} \bigg\{ F(r) \Big[ \frac{2}{r} +\frac{r}{3} \psi'(r)^2 \Big] +\frac{r}{3} \mu^2 \psi(r)^2 +\frac{2}{r} -\frac{4r}{L^2} \bigg\} =0\,,
\notag\\
&\chi' (r)-\frac{r}{3 [1-V(\lambda,r)]} \psi'(r)^2=0\,,
\label{eq.3.2.2}
\end{align}
where we have defined, 
$V(\lambda,r)=\frac{2\lambda L^2}{r^2} \big(1+F(r) \big)$.
Since in the Gauss-Bonnet gravity the equations of motion remain the second order differential equations, the number of boundary conditions are the same as in Einstein gravity. Similarly to the previous section, we need to know the behavior of every unknown function near the horizon $r=r_H$ and near the boundary when $r \to \infty$. 

As always, the definition of the location of horizon $F(r=r_H)=0$ gives the first boundary condition on $F(r)$. On the other hand, as we mentioned previously, the asymptotic behavior of this function is given by
\begin{align}
F(r \to \infty) \sim \frac{r^2}{L^2_{\text{eff}}} -1\,, \qquad L^2_{\text{eff}} =\frac{2 \lambda L^2}{1- \sqrt{1-4\lambda}}\,.
\end{align}
We also demand that the scalar field is a regular function at the horizon so the $\psi (r_H)=\mathcal{O} (1)$.
If we consider the scalar field as a quantum field living in the curved space-time outside the black hole horizon, even in the presence of the Gauss-Bonnet terms, it should satisfy the Klein-Gordon equation. The solution of this equation decays at infinity as $\psi (r) \sim \frac{A_{(+)}}{r^{\Delta_+}} +\frac{A_{(-)}}{r^{\Delta_-}}$. 
Again, $A_{(\pm)}$ are the expectation values of the conformal operators with $\Delta_{\pm}$ conformal dimensions. 

To find a relation for $\Delta$, one needs to substitute
$F(r)= \frac{r^2}{L^2_{\text{eff}}} -1$
and $\psi (r) =A\, r^{-\Delta}$ into the equation of motion for the scalar field (the first equation in \eqref{eq.3.2.2}) in the limit of $r \to \infty$. In this regard, we will find the following values for the conformal dimension
\begin{align}
\Delta_\pm = 2+\sqrt{4+\mu^2 L_{\text{eff}}^2}\,.
\label{eq.3.2.5}
\end{align}
In what follows, we again assume the Dirichlet boundary conditions, $A_{(-)}=0$, therefore we expect  $\psi (r \to \infty) \sim \frac{A}{r^{\Delta_+}}$ asymptotically. 
It is also easy to derive the boundary conditions for $\chi (r)$ just by looking at the last equation of motion in \eqref{eq.3.2.2}
\begin{align}
\chi (r_H)=\mathcal{O} (1)\,, \qquad \chi (r \to \infty) =\mathcal{O} (r^{-2\Delta_+}) +\cdots
\end{align}
We can use the advantage of working  with the dimensionless parameters  which is desirable in the numerical methods.
To do this we change the radial coordinate from $r$ to $z$ via $z=\frac{r_H}{r}$  and  introduce a dimensionless mass parameter $m=-\mu^2 L^2$. We also use the scaling symmetry of the equations of motion to define, $z_0 =\frac{r_H}{L_{\text{eff}}}$. Then the equations of motion will be the following
\begin{align}
&\psi'' (z)+\frac{ \psi' (z)}{z[1-V(\lambda,z)]} \bigg\{1 +\frac{1}{F(z)} \Big[ 2-\frac{z_0^2}{f_\infty z^2}(4+\frac{m}{3 }\psi(z)^2)\Big]\!+\!V(\lambda,z) \bigg\} 
\!+\! \frac{\psi (z)}{F(z)} \frac{m z_0^2}{f_\infty z^4} =0,\notag \\
&F'(z) +\frac{1}{z[1-V(\lambda,z)]} \bigg\{- F(z) \Big[ 2 +\frac{z^2}{3} \psi'(z)^2 \Big] +\frac{z_0^2 m}{3z^2 f_\infty} \psi(z)^2 -2 +\frac{4z_0^2}{z^2 f_\infty} \bigg\}=0\,,\notag \\ 
&\chi'(z) +\frac{z}{3 [1-V(\lambda,z)]} \psi'(z)^2 =0\,,
\label{eq.3.2.6}
\end{align}
where 
\be
V(\lambda,z) =\frac{2\lambda z^2 f_\infty}{z_0^2} \big(1+F(z) \big)\,,\qquad
\frac{L^2}{L_\text{eff}^2} =f_\infty=\frac{1- \sqrt{1-4\lambda}}{2\lambda}\,.
\label{eq.3.2.7}
\ee
The boundary condition also can be written as
\begin{alignat}{3}
&F_{B} (z\to 0) \sim \frac{z_0^2}{z^2} -1, && \quad F_{H} (z \to 1) =0\,,\notag\\
&\psi_{B} (z \to 0) \sim C_+ z^{\Delta_+}, &&\quad \psi_{H} (z \to 1)=\mathcal{O} (1)\,,
\label{eq.3.2.8}\\
&\chi_{B} (z \to 0) \sim \mathcal{O} (z^{2\Delta_+} ), && \quad \chi_{H} (z \to 1) =\mathcal{O}(1)\,.\notag
\end{alignat}
These boundary conditions are identical to those of \eqref{BDCz}.

As we discussed in section two, in order to find a numerical solution for equations of motion, we need to compute the expansions of the metric and the scalar functions near the horizon and boundary. Once again, we substitute the near horizon expansion 
\eqref{eq.2.2.12} into the first two equations of motion in  \eqref{eq.3.2.6} and expand them near the $z=1$. We find
\begin{align}
\psi_{H} (z)&=\psi_1 +(z-1) \Big[ \frac{3m \psi_1 (z_0^2 -2\lambda f_\infty)}{z_0^2 (12+m \psi_1^2) -6f_\infty} \Big]+\cdots \notag \\
F_{H} (z)&=(z-1)\frac{z_0^2}{3 f_\infty} \Big[ \frac{6 f_\infty -z_0^2 (12 +m \psi_1^2)}{z_0^2 -2\lambda f_\infty} \Big]+\cdots
\label{eq.3.2.9}
\end{align}
 Similarly, we use the near boundary expansion at $z=0$ in the equation (\ref{NBE}), but here we note that the leading order is given by the value of $\Delta_{+}$ in the equation (\ref{eq.3.2.5}) or $\Delta_+= 2+\sqrt{4- \frac{m}{f_\infty}}$ equivalently. 

 Here we have a two-parameter family of the solutions which are controlled by the values of $m$ and $f_\infty$ or equivalently by $\mu$ and $\l$. In our numerical method, it would be simpler to fix the value of $\Delta_{+}$ firstly and then choose some appropriate values for the dimensionless mass and coupling.

Once again, in shooting method we first read the initial values of the fields and their derivatives near the horizon and then solve the equations of motion numerically, to find the values of the fields near the boundary.
By comparing with the values of the near boundary expansion we will find four equations for four unknown parameters $C_{+}, C_m, \psi_1$ and $z_0$. 
The values of these parameters can be used to compute the various physical and thermodynamical quantities of our problem. 

The temperature of the modified hairy black hole is given by $
\tilde T =\frac{e^{\chi (z)}}{4\pi z_0} |F'(z)|\Big|_{z=1}$, where
from equations \eqref{eq.3.2.9} and \eqref{eq.3.2.6} we find
\begin{align}
F'(1)&=\frac{z_0^2}{3 f_\infty} \Big( \frac{6 f_\infty -z_0^2 (12 +m \psi_1^2)}{z_0^2 -2\lambda f_\infty} \Big)\,,
\notag \\
\chi (1)&=\int_{z=\epsilon}^{1-\epsilon} \frac{-z}{3 [1-V(\lambda,z)]} \psi'(z)^2\, dz\,.
\end{align}
The next physical quantity will be the thermal entropy, which we have calculated  for the EGB black hole from the Wald's entropy formula in \eqref{eq.3.1.21}.
For the modified hairy black hole we can use again this equation just by replacing  $r_H\rightarrow z_0 L_{\text{eff}}$.
Hence
\begin{align}
S_{\text{therm}}= \frac{V_\Sigma L^3}{G_N} \frac{z_0^3}{4 f_\infty^{3/2}} \Big( 1-\frac{6\lambda f_\infty}{z_0^2} \Big) \equiv \Big( \frac{V_\Sigma L^3}{G_N} \Big) \tilde S\,.\label{sfi}
\end{align}
The computation of the energy is straightforward and we can use the same formula that we found in equations (\ref{eq.2.28}) and (\ref{eq.2.29}) again
\begin{align}
E=\frac{3 V_\Sigma}{16 \pi G_N} \tilde m z_0^2 L_{eff}^2\equiv \frac{V_\Sigma L^2}{G_N} \tilde E\,.
\end{align}
Since we are interested in the thermal phase transition, we need to find the condensate or the $\langle O \rangle= C_+ (\frac{z_0}{\sqrt{f_\infty}})^{\Delta_{+}}$.
By plotting the condensate versus the temperature (see figure \ref{fig6}), we observe that the scalar field condensates at every value of the $\lambda$ within the allowed interval
$-\frac{7}{36} \le \lambda \le \frac{9}{100}$.
As we see in figure \ref{fig6}, for the fixed value of $\Delta_{+}=2$, the critical temperature decreases when the Gauss-Bonnet coupling increases.
\begin{figure}[!htbp]
\centering
\includegraphics[width=1\textwidth]{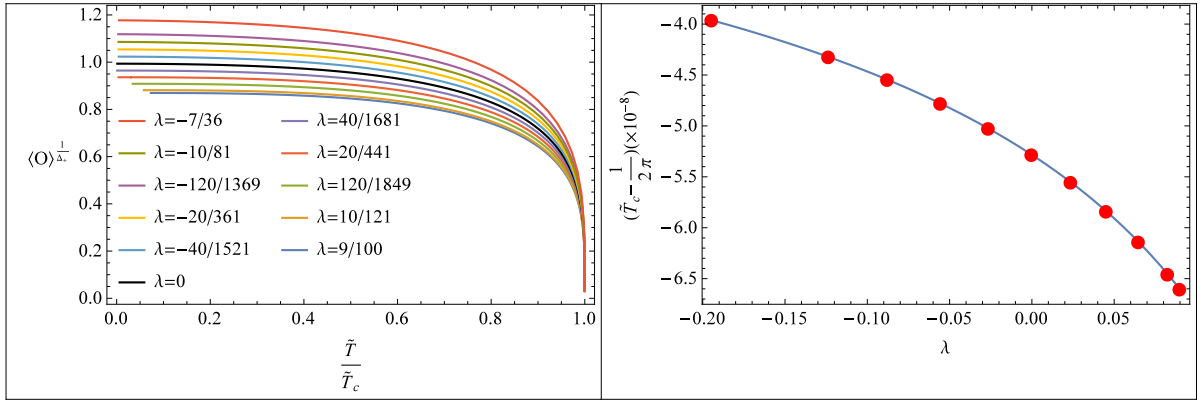}
\caption{\footnotesize{Left: The condensate as a function of the temperature (scaled by the corresponding critical temperature) for the various values of  $\lambda$ when $\Delta_+=2$. The lowest curve is for $\lambda=\frac{9}{100}$ and the top one corresponds to the $\lambda=-\frac{7}{36}$. Right: The critical temperature as a function of $\lambda$.}}
\label{fig6}
\end{figure}

To see the relationship between the various thermodynamical quantities and moreover to follow how the condensation happens in this theory, we have depicted their graphs in figure \ref{fig7}. 
\begin{figure}[!htbp]
	\centering
	\includegraphics[width=0.95\textwidth]{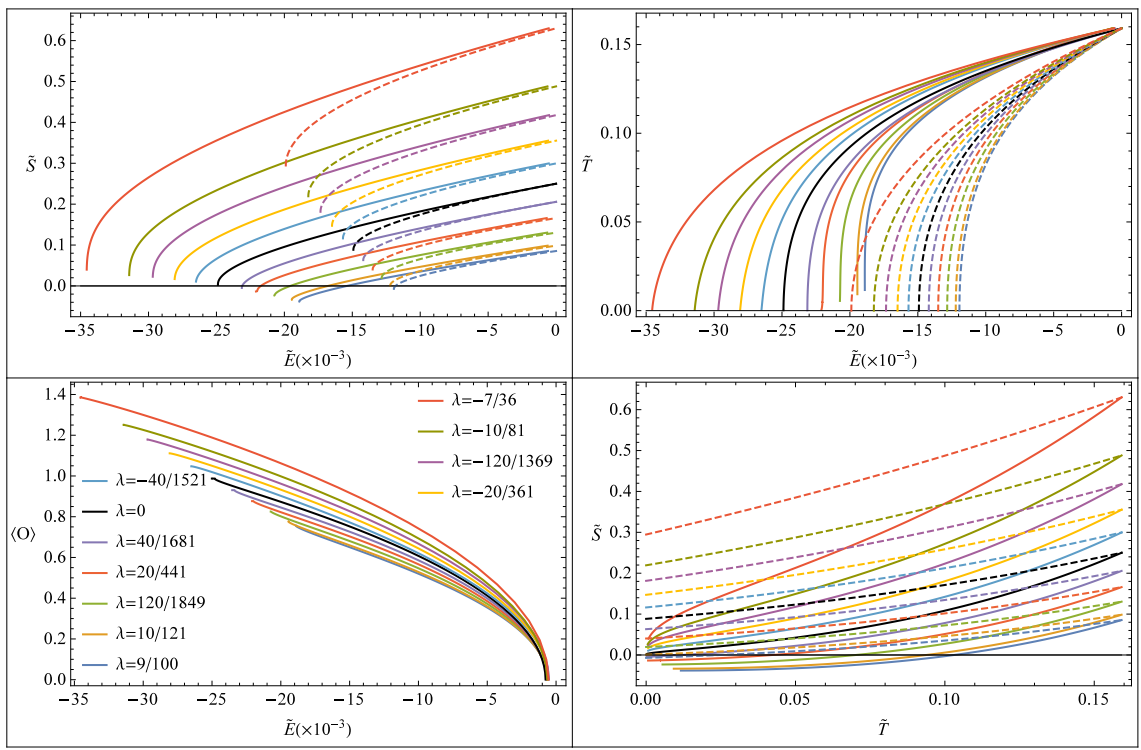}
	\caption{\footnotesize{The thermodynamical quantities for the various values of $\lambda$ when $\Delta_+=2$. The colors in these graphs are matched with those in figure \ref{fig6}. In each graph, a horizontal black line is sketched to guide the zero temperature and zero thermal entropy. The dashed curves represent the EGB black hole, while the solid ones represent the modified hairy black holes.}}
	\label{fig7}
\end{figure}
The behavior of various quantities is the same as in the case of Einstein gravity. We can see this by comparing the curves of $\l\neq 0$ with the curve $\l=0$ (black curve) in each graph. For each value of the coupling, there is a critical temperature where a phase transition from the EGB black hole to the modified hairy black hole happens. 

But there is an important point which can be seen easily in the $\tilde S-\tilde E$ or the $\tilde S-\tilde T$ diagrams. For the positive values of the coupling $\l$, there is a temperature or energy below the critical point, where the value of the entropy becomes negative. The value of these temperatures, $T_v$, is depicted in figure \ref{fig8}. The same behavior is happening for the EGB thermal entropy in a narrow region of the couplings $\frac{1}{12}<\l<\frac{9}{100}$, this has been already reported in \cite{Pastras:2014oka} and \cite{Pastras:2015mza}.
\begin{figure}[!htbp]
	\centering
	\includegraphics[width=.46\textwidth]{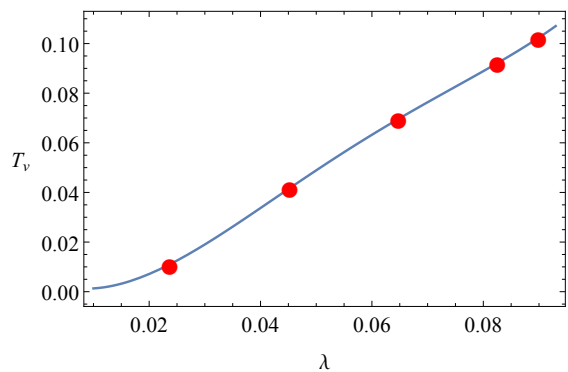}
	\caption{\footnotesize{For each positive value of the $\l$ there is a temperature, $T_v$, where the thermal entropy becomes negative.}}
	\label{fig8}
\end{figure}

Finally the R\'enyi entropies should be computed using  the EGB temperature in the equation \eqref{eq.3.1.8}. By considering the dimensionless temperature, $\tilde X_c$ will be the real positive root of the following equation
\begin{align}
\tilde X_c^3 -\tilde X_c^2 (\pi \tilde T_c f_\infty) -\tilde X_c \Big( \frac{f_\infty}{2} \Big) +2\pi \lambda \tilde T_c f_\infty^2 =0\,.
\label{eq.3.2.17}
\end{align}
Then according to  \eqref{eq.3.1.23}, the dimensionless entanglement R\'enyi entropies  yield
\begin{align}
\tilde S_n &= \frac{2\pi n}{n-1} \int_{\tilde T_0/n}^{\tilde T_c} \tilde{S}_\text{thermal}^\text{HBH} (\tilde T) \, d\tilde T +\Big( \frac{n}{n-1} \Big) \Big( \frac{1}{8f_\infty^{5/2}} \Big) \bigg\{ 9(1-\tilde X_c^4 )-3 f_\infty (1-\tilde X_c^2) \notag\\& 
+\frac{4( f_\infty -2)}{1-2\lambda f_\infty}-
\frac{4 \tilde X_c^4 (f_\infty -2\tilde X_c^2)}{\tilde X_c^2 -2\lambda f_\infty} \bigg\}\,,\label{SHBH}
\end{align}
where $\tilde T_0=\frac{1}{2\pi}$.
Once again the first term of this expression should be computed numerically. Since $\tilde T_c\geq \frac{\tilde T_0}{n}$  we can define a critical value, $n_c=\frac{\tilde T_0}{\tilde T_c}$, where the R\'enyi entropy corresponding to the modified hairy black hole approaches that of the EGB black hole, see figure \ref{fig9}. 
\begin{figure}[!htbp]
	\centering
	\includegraphics[width=1\textwidth]{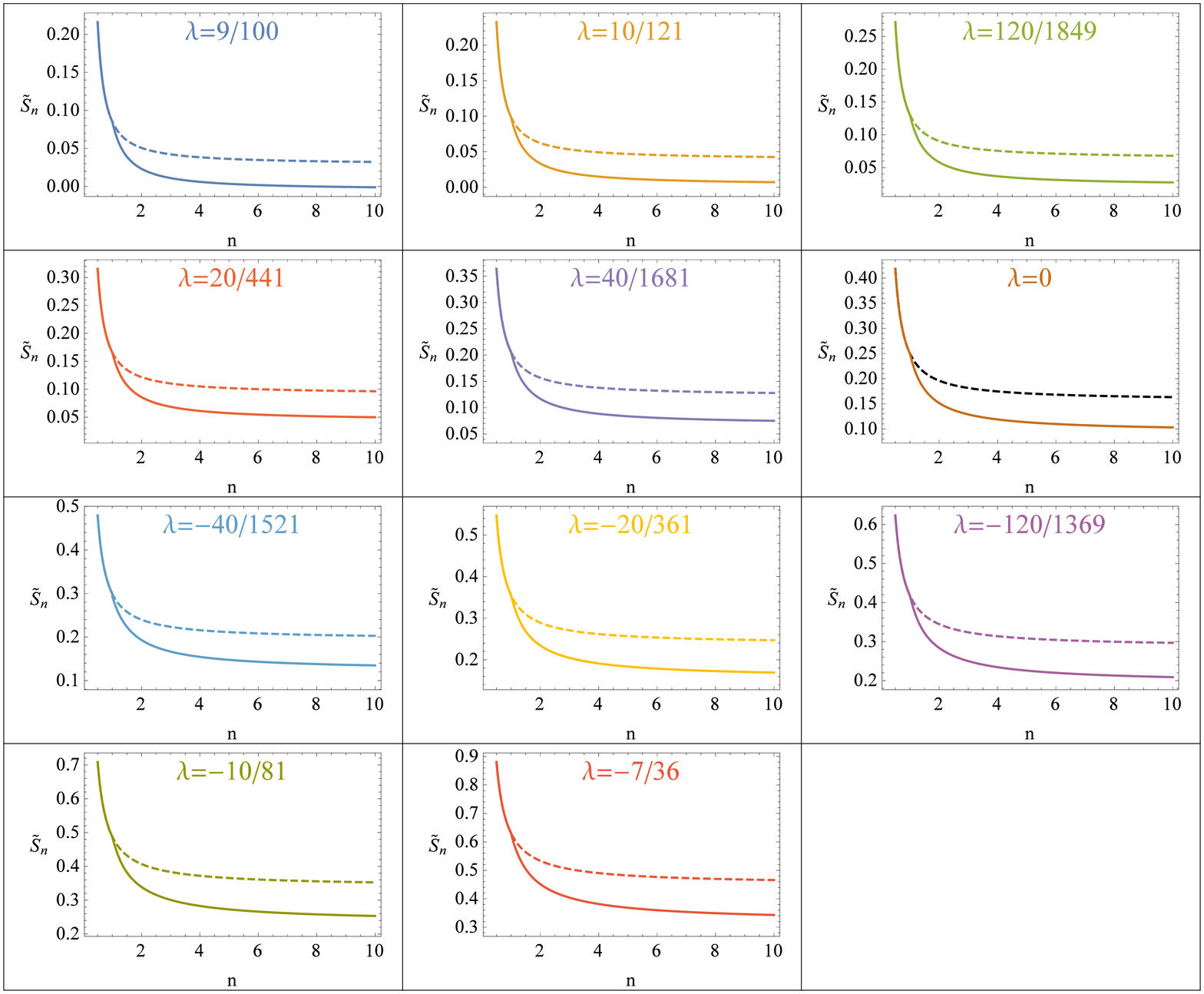}
	\caption{\footnotesize{In each diagram there is a critical point where the ERE of the modified hairy black hole (solid curve) approaches that of the EGB black hole (dashed curve).
	}}
	\label{fig9}
\end{figure}
These diagrams show that in the presence of the Gauss-Bonnet terms the condensation of the scalar field is again a second order phase transition.

To show the behavior of $n_c$ under the change of the coupling $\l$, we have sketched the figure \ref{fig10} (left). As we see, this critical value increases by moving from the negative values of the coupling to the positive values.
\begin{figure}[!htbp]
\centering
\includegraphics[width=1\textwidth]{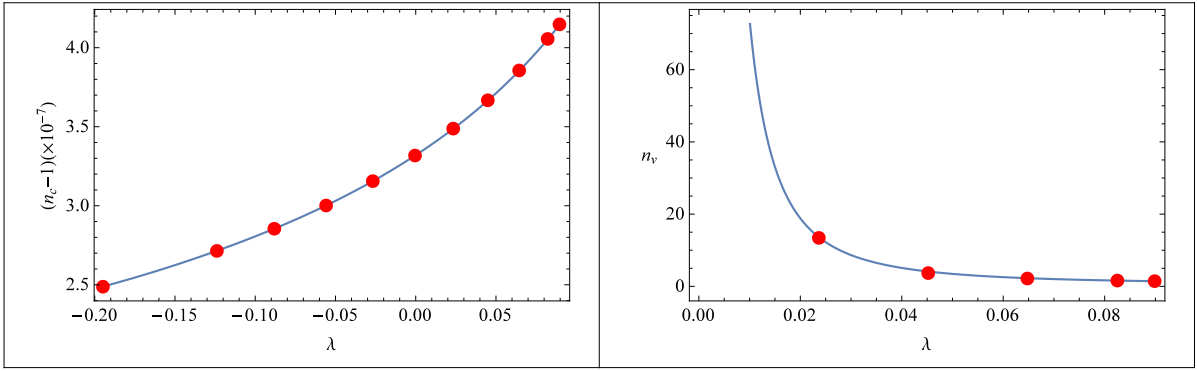}
\caption{\footnotesize{Left: When the scalar mass is $\Delta_+=2$, the critical R\'enyi parameters are extremely close to one. So we have sketched $n_c -1$ as a function of $\lambda$. Right: The $n_v$ shows the value of $n$, where the second inequality governing the EREs is violated. This diagram shows the values of $n_v$ as a function of $\lambda$,  for all the positive values of the GB coupling constant.}}
\label{fig10}
\end{figure}

Since we have the numerical behavior of the R\'enyi entropies, it is worth checking the inequalities of (\ref{1NEQ})-(\ref{4NEQ}).  We have sketched the corresponding diagrams in terms of the various values of $n$, see figure \ref{in1234}.  

\begin{figure}[!htbp]
\centering
\includegraphics[width=0.95\textwidth]{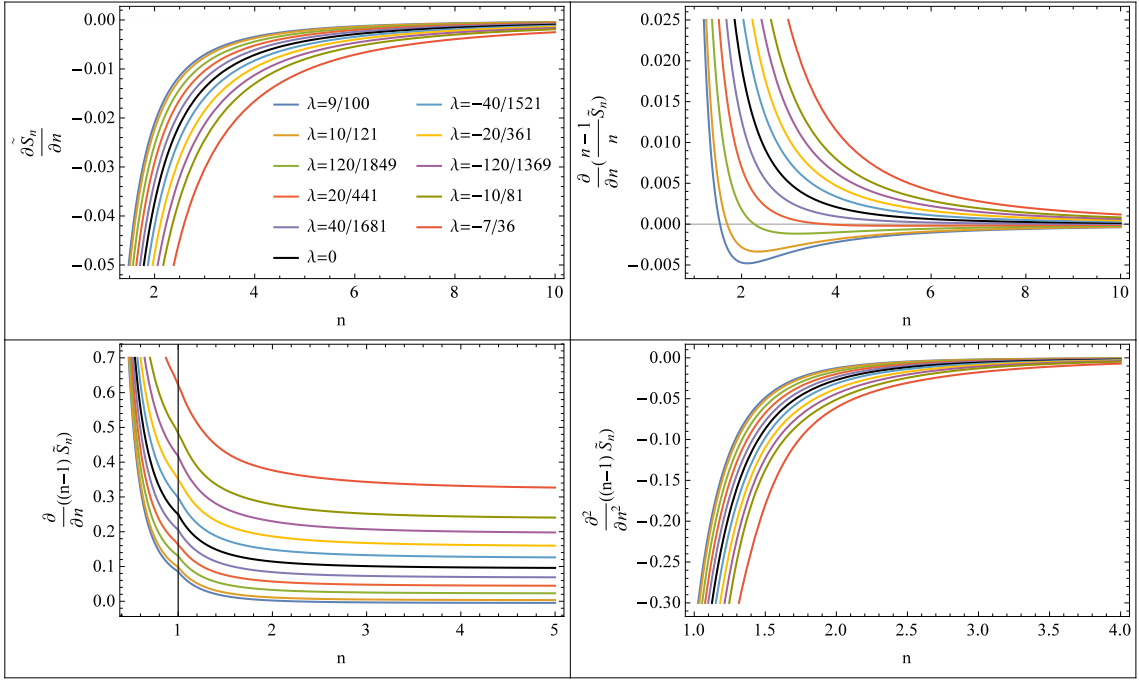}
\caption{Inequalities of R\'enyi entropies. The (down-left) diagram shows that the phase transition is a second order one. }\label{in1234}
\end{figure}

As we observe, the inequalities (\ref{1NEQ}) (up-left)    and (\ref{4NEQ}) (down-right) hold here but the second inequality (\ref{2NEQ}) (up-right) is violated for the positive values of the coupling $\l$ (there is an interval for $n$, where the curves get the negative values). This is completely consistent with the fact that for the positive values of the coupling, there is a temperature where the thermal entropy becomes negative. 
By a numerical study, one can find a violation number, $n_v$,  where the second inequality governing the EREs is violated, see the right diagram in figure \ref{fig10}.
The third inequality (\ref{3NEQ}) (down-left) also holds here, and as we see, there is a discontinuity in the slope of the curves at $n=1$. This shows a second order phase transition. We will discuss these results in the next section.

\subsubsection{Results for \texorpdfstring{$\Delta_{+} > 2$}{TEXT}}
To complete our results, we have drawn the figure \ref{fig12} for other values of $\Delta_{+} > 2$. In this figure we have presented the critical temperature $\tilde T_c$, the critical R\'enyi parameter $n_c$ and the violation number $n_v$ for each value of the Gauss-Bonnet coupling $\l$. The behavior of these parameters are similar to those of $\Delta_{+}=2$ but is more pronounced as $\Delta_{+}$ increases.
\begin{figure}[!htbp]
\centering
\includegraphics[width=1\textwidth]{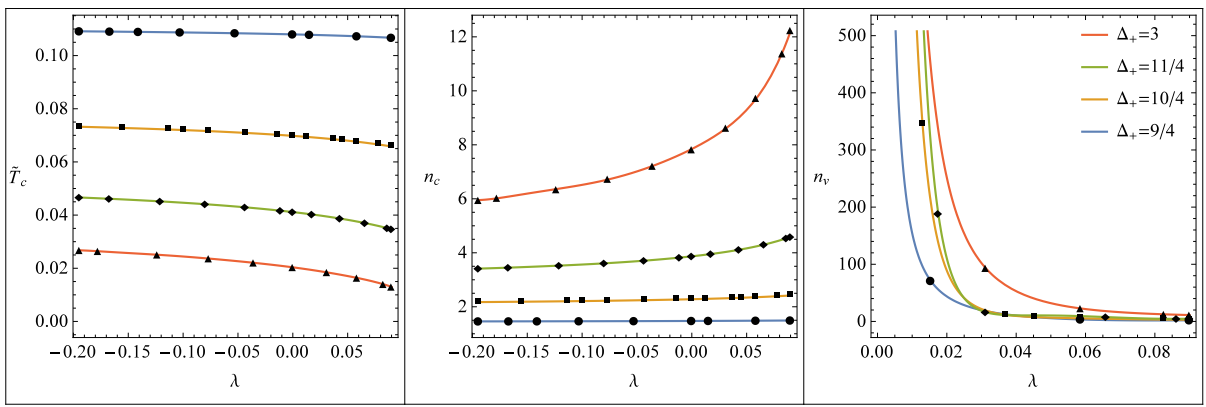}
\caption{\footnotesize{From left to right, the critical temperature, the critical R\'enyi parameter, and the violation number in terms of the GB coupling}}
\label{fig12}
\end{figure}
\section{Discussion}
In section 2, we have reviewed the phase transition from the Einstein black hole to a hairy black hole due to the condensation of a scalar field. Both solutions exist in a five-dimensional space-time and have a hyperbolic spatial boundary. The numerical results of the condensate in terms of the temperature are sketched in figure \ref{fig2} (the left diagram). We have also found the critical temperatures for this phase transition within the unitarity bound $2\leq\Delta_{+}\leq 2+\sqrt{3}$, see figure \ref{fig2} (the right diagram). These diagrams show that, by increasing the value of the conformal dimension of the scalar field, the critical temperature of the phase transition decreases. Moreover, the value of the condensate decreases too.

One can explain this condensation in terms of the energy, by finding the relation between the energy of the black holes and their temperatures, via the $\tilde{T}-\tilde{E}$ diagram in figure \ref{fig3}.
The  $\tilde{S}-\tilde{E}$ diagram shows that the entropy of a hairy black hole at a specific value of the energy is larger than the entropy of the Einstein black hole, so the hairy black hole is prefered to the Einstein black hole after the condensation. The $\langle O\rangle-\tilde{E}$ curves display a phase transition when the energy decreases.

As we reviewed in the introduction, by knowing the thermal entropy, one can compute the entanglement R\'enyi entropies $S_n$, for a spherical region in the dual gauge theory. The results for these entropies are given in the $\tilde S_n-n$ diagrams of figure \ref{fig4}. They show the existence of a critical value, $n_c$, where a discontinuity is happening for the second derivative of the $S_n$. This suggests that the condensation is a second order phase transition.

In section 3 we have extended the study of the phase transition in section 2, by adding the Gauss-Bonnet terms to the action. We have found the modified black hole solutions before and after the condensation, which now depend on the GB coupling $\l$ in the allowed region $-\frac{7}{36} \le \lambda \le \frac{9}{100}$. The effect of adding this coupling on the condensation is presented in figures \ref{fig6} and \ref{fig7}. 
Here we have two important parameters, the conformal dimension of the scalar field, $\Delta_+$ and the coupling $\l$. We have presented the results for $\Delta_+=2$ completely and for $\Delta_+>2$ briefly in this paper.
 
We show that by going from the positive values of the coupling to the negative ones the value of the condensate and the critical temperature are increasing, see figure \ref{fig6}.
By looking at the diagrams  $\tilde S-\tilde E$ or $\tilde S-\tilde T$ in figure \ref{fig7}, we observe a new behavior for $\l> 0$. For the positive values of the coupling $\l$, there is a temperature or energy, below the critical point, where the value of the entropy becomes negative. The value of these temperatures, $T_v$, is depicted in figure \ref{fig8}. 

As it was mentioned, the violation of the second inequality of the R\'enyi  entropy (\ref{2NEQ}), is related to the negativity of the black hole entropy. In references \cite{Pastras:2014oka} and \cite{Pastras:2015mza}, the authors have studied the same violation in the Gauss-Bonnet gravitational theory in the presence of a gauge field and found a new bound on the coupling of the theory $i.e.$ $-\frac{7}{36} \le \lambda \le \frac{1}{12}$ in 5 dimensions, instead of the interval demanded by the unitarity of the theory $i.e.$ $-\frac{7}{36} \le \lambda \le \frac{9}{100}$. In this paper, we show that for the Gauss-Bonnet gravity and in the presence of a scalar field, if we demand the positivity of the black hole thermal entropy, then there will be a novel bound on the coupling of the theory $i.e.$ $-\frac{7}{36} \le \lambda \le 0$.

The entanglement R\'enyi  entropy is calculated in the equation (\ref{SHBH}) and for the various values of the coupling are depicted in figure \ref{fig9}. As we see, a second order phase transition is happening again. The critical points of $n_c$, where the ERE  of the hairy black holes meet the ERE of the EGB black holes, have been drawn in the left diagram of the figure \ref{fig10}.

One of the important subjects in the study of the entanglement R\'enyi  entropy is the existence of the inequalities. We have checked some of these inequalities, (\ref{1NEQ})-(\ref{4NEQ}), and the reader can see the results of our numerical computations in figure \ref{in1234}. 

Let us explain our results according to the equation (\ref{SHBH}) and write it in a little simpler form
\begin{align}
\tilde S_n=\frac{2\pi n}{n-1} \int_{\tilde T_0/n}^{\tilde T_c} S^{\text{HBH}}(\tilde T) d\tilde T+ \frac{n}{n-1} \xi\,,
\end{align}
where $\xi$ is a positive number independent of $n$.
As it was discussed, the inequalities (\ref{1NEQ}) and (\ref{4NEQ}) hold in our case. To explain the first inequality we can write
\begin{align}
\frac{\partial }{\partial n}\tilde S_n=-\frac{2\pi}{(n-1)^2}
\int_{\tilde T_0/n}^{\tilde T_c}\Big[S^{\text{HBH}}(\tilde T)-S^{\text{HBH}}(\frac{\tilde T_0}{n})\Big] d\tilde T-\frac{1}{(n-1)^2}\xi\,.
\end{align}
The first term in the above equation is negative, because the $S^{\text{HBH}}$ is a monotonically increasing function of the temperature, see the $\tilde S-\tilde T$ diagram in the figure \ref{fig7}. Since the $\xi$ is a positive constant, the total value of the right hand side of the above expression will be negative.

The fourth inequality can be written in terms of the specific heat, to see this we notice that
\begin{align}
\frac{\partial^2 }{\partial n^2}\big((n-1)\tilde S_n\big)=-\frac{1}{n^2}\frac{\partial \tilde E}{\partial \tilde T}\Big|_{\tilde T_0/n}\,.
\end{align}
The $\tilde T-\tilde E$ diagram in figure \ref{fig7} easily shows that the right hand side of the above equation is always negative. 
 
Now let us look at the second inequality (\ref{2NEQ}), which is violated for the positive values of the coupling $\l$, as depicted in the figure \ref{in1234}. This is completely consistent with the fact that for the positive values of the coupling there is a temperature where the thermal entropy becomes negative. In fact, we have
\begin{align}
\frac{\partial }{\partial n}\big(\frac{n-1}{n}\tilde S_n\big)=\frac{1}{n^2} S^{\text{HBH}}(\frac{\tilde T_0}{n})\,.
\end{align}
The $\tilde S-\tilde T$ diagram in the figure \ref{fig7} proves that whenever the $\frac{\tilde T_0}{n}$ is smaller than a specific temperature then the value of the thermal entropy becomes negative and therefore the second inequality is violated. We have sketched the violation number $n_v$, in the right diagram of the figure \ref{fig12} for the positive values of the coupling. In fact, this violation never occurs for $\l\leq 0$, because as we see from equation (\ref{sfi}), for the negative values of the coupling, the thermal entropy is always positive.

The third inequality (\ref{3NEQ}) has a more complicated form
\begin{align}
\frac{\partial }{\partial n}\big((n-1)\tilde S_n\big)=2\pi
\int_{\tilde T_0/n}^{\tilde T_c} S^{\text{HBH}}(\tilde T) d\tilde T +\frac{1}{n} \tilde S^{\text{HBH}}(\frac{\tilde T_0}{n})+\xi\,.
\end{align}
The first and the second terms above, may have  positive or  negative values, depending on the value of the $\frac{\tilde T_0}{n}$. The last term is always positive, therefore it would be difficult to find the total sign of the above expression, otherwise we use the numerical data. Our numerical analysis shows that, this inequality holds for the most areas of the parameters of the theory as depicted in the figure \ref{in1234} (down-left). 

By more numerical analysis for the other values of the $\Delta_{+}>2$, we conclude that all the above arguments and discussions are $\Delta_{+}$ independent, see figure \ref{fig12}.
\section*{Acknowledgment}
This work is supported by Ferdowsi University of Mashhad under the grant 3/42540 (1395/10/28). A. G. would like to thank M. Roshan and K. Javidan for valuable discussions on the numerical methods. S. Q. would like to thank  J. E. Santos for useful discussions.


\begin{thebibliography}{99}

\bibitem{Abramsky:2003}
  S.~Abramsky and B.~Coecke,
  {\it Physical Traces: Quantum vs. Classical Information Processing},
[arXiv:cs/0207057v2].

\bibitem{Calabrese:2004eu} 
  P.~Calabrese and J.~L.~Cardy,
  ``Entanglement entropy and quantum field theory,''
  J.\ Stat.\ Mech.\  {\bf 0406}, P06002 (2004)
  [hep-th/0405152].

\bibitem{Hamma:2005}
  A.~Hamma, R.~Ionicioiu and P.~Zanardi,
{\it Ground State Entanglement and Geometric Entropy in the Kitaev's Model},
Phys.\ Lett.\ A {\bf 337} (2005) 22
[arXiv:quant-ph/0406202].
  
\bibitem{Calabrese:2005zw} 
  P.~Calabrese and J.~L.~Cardy,
  ``Entanglement entropy and quantum field theory: A Non-technical introduction,''
  Int.\ J.\ Quant.\ Inf.\  {\bf 4}, 429 (2006)
  [quant-ph/0505193].

\bibitem{Kitaev:2005dm} 
  A.~Kitaev and J.~Preskill,
  ``Topological entanglement entropy,''
  Phys.\ Rev.\ Lett.\  {\bf 96}, 110404 (2006)
  [hep-th/0510092].
  
\bibitem{Ryu:2006bv} 
  S.~Ryu and T.~Takayanagi,
  ``Holographic derivation of entanglement entropy from AdS/CFT,''
  Phys.\ Rev.\ Lett.\  {\bf 96}, 181602 (2006)
  [hep-th/0603001].

\bibitem{Ryu:2006ef} 
  S.~Ryu and T.~Takayanagi,
  ``Aspects of Holographic Entanglement Entropy,''
  JHEP {\bf 0608}, 045 (2006)
  [hep-th/0605073].

\bibitem{Nishioka:2009un} 
  T.~Nishioka, S.~Ryu and T.~Takayanagi,
  ``Holographic Entanglement Entropy: An Overview,''
  J.\ Phys.\ A {\bf 42}, 504008 (2009)
  [arXiv:0905.0932 [hep-th]].

\bibitem{VanRaamsdonk:2009ar} 
  M.~Van Raamsdonk,
  ``Comments on quantum gravity and entanglement,''
 [arXiv: 0907.2939 [hep-th]].
 
\bibitem{VanRaamsdonk:2010pw} 
  M.~Van Raamsdonk,
  ``Building up spacetime with quantum entanglement,''
  Gen.\ Rel.\ Grav.\  {\bf 42}, 2323 (2010)
  [Int.\ J.\ Mod.\ Phys.\ D {\bf 19}, 2429 (2010)]
  [arXiv:1005.3035 [hep-th]].


\bibitem{Takayanagi:2012kg} 
  T.~Takayanagi,
  ``Entanglement Entropy from a Holographic Viewpoint,''
  Class.\ Quant.\ Grav.\  {\bf 29}, 153001 (2012)
  [arXiv:1204.2450 [gr-qc]].

\bibitem{Bianchi:2012ev} 
  E.~Bianchi and R.~C.~Myers,
  ``On the Architecture of Spacetime Geometry,''
  Class.\ Quant.\ Grav.\  {\bf 31}, 214002 (2014)
  [arXiv:1212.5183 [hep-th]].

\bibitem{Myers:2013lva} 
  R.~C.~Myers, R.~Pourhasan and M.~Smolkin,
  ``On Spacetime Entanglement,''
  JHEP {\bf 1306}, 013 (2013)
  [arXiv:1304.2030 [hep-th]].

\bibitem{Balasubramanian:2013rqa} 
  V.~Balasubramanian, B.~Czech, B.~D.~Chowdhury and J.~de Boer,
  ``The entropy of a hole in spacetime,''
  JHEP {\bf 1310}, 220 (2013)
  [arXiv:1305.0856 [hep-th]].

\bibitem 
 {Renyi0} A.~R\'enyi, ``On measures of information and entropy,'' in {\sl Proceedings of the
4th Berkeley Symposium on Mathematics, Statistics and Probability},
{\bf 1}, 547 (U. of California Press, Berkeley, CA, 1961);
A.~R\'enyi, ``On the foundations of information theory,'' Rev.\ Int.\ Stat.\ Inst.\ {\bf 33} (1965) 1.

\bibitem{Li-Haldane}
H.~ Li and F.~D.~M.~ Haldane,
``Entanglement Spectrum as a Generalization of Entanglement Entropy: Identification of Topological Order in Non-Abelian Fractional Quantum Hall Effect States"
Phys. Rev. Lett. 101, 010504 (2008) [arXiv:0805.0332]

\bibitem{Cardy:2007mb} 
  J.~L.~Cardy, O.~A.~Castro-Alvaredo and B.~Doyon,
  ``Form factors of branch-point twist fields in quantum integrable models and entanglement entropy,''
  J.\ Statist.\ Phys.\  {\bf 130}, 129 (2008)
  [arXiv:0706.3384 [hep-th]].

\bibitem{Nishioka:2006gr} 
  T.~ Nishioka and T.~ Takayanagi,
  ``AdS Bubbles, Entropy and Closed String Tachyons,''
  JHEP {\bf 0701}, 090 (2007)
  [hep-th/0611035].
  
\bibitem{Velytsky:2008rs} 
  A.~Velytsky,
  ``Entanglement entropy in d+1 SU(N) gauge theory,''
  Phys.\ Rev.\ D {\bf 77}, 085021 (2008)
  [arXiv:0801.4111 [hep-th]].

\bibitem{Buividovich:2008kq} 
  P.~V.~Buividovich and M.~I.~Polikarpov,
  ``Numerical study of entanglement entropy in SU(2) lattice gauge theory,''
  Nucl.\ Phys.\ B {\bf 802}, 458 (2008)
  [arXiv:0802.4247 [hep-lat]].

\bibitem{Lewkowycz:2013nqa} 
  A.~Lewkowycz and J.~Maldacena,
  ``Generalized gravitational entropy,''
  JHEP {\bf 1308}, 090 (2013)
  [arXiv:1304.4926 [hep-th]].

\bibitem{Casini:2011kv} 
  H.~Casini, M.~Huerta and R.~C.~Myers,
  ``Towards a derivation of holographic entanglement entropy,''
  JHEP {\bf 1105}, 036 (2011)
  [arXiv:1102.0440 [hep-th]].

\bibitem{Aminneborg:1996iz} 
  S.~Aminneborg, I.~Bengtsson, S.~Holst and P.~Peldan,
  ``Making anti-de Sitter black holes,''
  Class.\ Quant.\ Grav.\  {\bf 13}, 2707 (1996)
  [gr-qc/9604005].

\bibitem{Brill:1997mf} 
  D.~R.~Brill, J.~Louko and P.~Peldan,
  ``Thermodynamics of (3+1)-dimensional black holes with toroidal or higher genus horizons,''
  Phys.\ Rev.\ D {\bf 56}, 3600 (1997)
  [gr-qc/9705012].

\bibitem{Vanzo:1997gw} 
  L.~Vanzo,
  ``Black holes with unusual topology,''
  Phys.\ Rev.\ D {\bf 56}, 6475 (1997)
  [gr-qc/9705004].

\bibitem{Mann:1996gj} 
  R.~B.~Mann,
  ``Pair production of topological anti-de Sitter black holes,''
  Class.\ Quant.\ Grav.\  {\bf 14}, L109 (1997)
  [gr-qc/9607071].

\bibitem{Birmingham:1998nr} 
  D.~Birmingham,
  ``Topological black holes in Anti-de Sitter space,''
  Class.\ Quant.\ Grav.\  {\bf 16}, 1197 (1999)
  [hep-th/9808032].

\bibitem{Emparan:1998he} 
  R.~Emparan,
  ``AdS membranes wrapped on surfaces of arbitrary genus,''
  Phys.\ Lett.\ B {\bf 432}, 74 (1998)
  [hep-th/9804031].

\bibitem{Wald:1993nt} 
  R.~M.~Wald,
  ``Black hole entropy is the Noether charge,''
  Phys.\ Rev.\ D {\bf 48}, no. 8, R3427 (1993)
  [gr-qc/9307038].

\bibitem{Iyer:1994ys} 
  V.~Iyer and R.~M.~Wald,
  ``Some properties of Noether charge and a proposal for dynamical black hole entropy,''
  Phys.\ Rev.\ D {\bf 50}, 846 (1994)
  [gr-qc/9403028].

\bibitem{Hung:2011nu} 
  L.~Y.~Hung, R.~C.~Myers, M.~Smolkin and A.~Yale,
  ``Holographic Calculations of Renyi Entropy,''
  JHEP {\bf 1112}, 047 (2011)
  [arXiv:1110.1084 [hep-th]].

\bibitem{Caputa:2013eka} 
  P.~Caputa, G.~Mandal and R.~Sinha,
  ``Dynamical entanglement entropy with angular momentum and U(1) charge,''
  JHEP {\bf 1311}, 052 (2013)
  [arXiv:1306.4974 [hep-th]].

\bibitem{Belin:2013uta} 
  A.~Belin, L.~Y.~Hung, A.~Maloney, S.~Matsuura, R.~C.~Myers and T.~Sierens,
  ``Holographic Charged Renyi Entropies,''
  JHEP {\bf 1312}, 059 (2013)
  [arXiv:1310.4180 [hep-th]].

\bibitem{Pastras:2014oka} 
  G.~Pastras and D.~Manolopoulos,
  ``Charged Rényi entropies in CFTs with Einstein-Gauss-Bonnet holographic duals,''
  JHEP {\bf 1411}, 007 (2014)
  [arXiv:1404.1309 [hep-th]].


\bibitem{Pastras:2015mza} 
  G.~Pastras and D.~Manolopoulos,
  ``Holographic Calculation of Renyi Entropies and Restrictions on Higher Derivative Terms,''
  PoS CORFU {\bf 2014}, 157 (2015)
  [arXiv:1507.08595 [hep-th]].

\bibitem{karol}
K.~Zyczkowski, ``R\'enyi extrapolation of Shannon entropy,'' Open Syst.
Inf. Dyn. {\bf 10}, 297 (2003) [arXiv:quant-ph/0305062].

\bibitem{Nakaguchi:2016zqi} 
  Y.~Nakaguchi and T.~Nishioka,
  ``A holographic proof of Rényi entropic inequalities,''
  JHEP {\bf 1612}, 129 (2016)
  [arXiv:1606.08443 [hep-th]].

\bibitem{Belin:2013dva} 
  A.~Belin, A.~Maloney and S.~Matsuura,
  ``Holographic Phases of Renyi Entropies,''
  JHEP {\bf 1312}, 050 (2013)
  [arXiv:1306.2640 [hep-th]].

\bibitem{Hartnoll:2008kx} 
  S.~A.~Hartnoll, C.~P.~Herzog and G.~T.~Horowitz,
  ``Holographic Superconductors,''
  JHEP {\bf 0812}, 015 (2008)
  [arXiv:0810.1563 [hep-th]].

\bibitem{Horowitz:2010gk} 
  G.~T.~Horowitz,
  ``Introduction to Holographic Superconductors,''
  Lect.\ Notes Phys.\  {\bf 828}, 313 (2011)
  [arXiv:1002.1722 [hep-th]].


\bibitem{Dias:2010ma} 
  O.~J.~C.~Dias, R.~Monteiro, H.~S.~Reall and J.~E.~Santos,
  ``A Scalar field condensation instability of rotating anti-de Sitter black holes,''
  JHEP {\bf 1011}, 036 (2010)
  [arXiv:1007.3745 [hep-th]].

\bibitem{Ashtekar:1999jx} 
A.~Ashtekar and S.~Das,
``Asymptotically Anti-de Sitter space-times: Conserved quantities,''
Class.\ Quant.\ Grav.\  {\bf 17}, L17 (2000)
[hep-th/9911230].


\bibitem{Breitenlohner:1982jf}
  P.~Breitenlohner and D.~Z.~Freedman,
  ``Stability In Gauged Extended Supergravity,''
  Annals Phys.\  {\bf 144} (1982) 249; \\
  P.~Breitenlohner and D.~Z.~Freedman,
  ``Positive Energy In Anti-De Sitter Backgrounds And Gauged Extended
  Supergravity,''
  Phys.\ Lett.\  B {\bf 115} (1982) 197.

\bibitem{Mezincescu:1984ev}
  L.~Mezincescu and P.~K.~Townsend,
  ``Stability At A Local Maximum In Higher Dimensional Anti-De Sitter Space And
  Applications To Supergravity,''
  Annals Phys.\  {\bf 160} (1985) 406.

\bibitem{Aros:2002te} 
  R.~Aros, C.~Martinez, R.~Troncoso and J.~Zanelli,
  ``Quasinormal modes for massless topological black holes,''
  Phys.\ Rev.\ D {\bf 67}, 044014 (2003)
  [hep-th/0211024].

\bibitem{Klebanov:1999tb} 
  I.~R.~Klebanov and E.~Witten,
  ``AdS / CFT correspondence and symmetry breaking,''
  Nucl.\ Phys.\ B {\bf 556}, 89 (1999)
  [hep-th/9905104].
	
\bibitem{Cai:2001dz} 
  R.~G.~Cai,
  ``Gauss-Bonnet black holes in AdS spaces,''
  Phys.\ Rev.\ D {\bf 65}, 084014 (2002)
  [hep-th/0109133].

\bibitem{Cvetic:2001bk} 
  M.~Cvetic, S.~Nojiri and S.~D.~Odintsov,
  ``Black hole thermodynamics and negative entropy in de Sitter and anti-de Sitter Einstein-Gauss-Bonnet gravity,''
  Nucl.\ Phys.\ B {\bf 628}, 295 (2002)
  [hep-th/0112045].

\bibitem{Anninos:2008sj} 
  D.~Anninos and G.~Pastras,
  ``Thermodynamics of the Maxwell-Gauss-Bonnet anti-de Sitter Black Hole with Higher Derivative Gauge Corrections,''
  JHEP {\bf 0907}, 030 (2009)
  [arXiv:0807.3478 [hep-th]].

\bibitem{Bhattacharjee:2015qaa} 
  S.~Bhattacharjee, A.~Bhattacharyya, S.~Sarkar and A.~Sinha,
  ``Entropy functionals and c-theorems from the second law,''
  Phys.\ Rev.\ D {\bf 93}, no. 10, 104045 (2016)
  [arXiv:1508.01658 [hep-th]].

\bibitem{Ge:2008ni} 
  X.~H.~Ge, Y.~Matsuo, F.~W.~Shu, S.~J.~Sin and T.~Tsukioka,
  ``Viscosity Bound, Causality Violation and Instability with Stringy Correction and Charge,''
  JHEP {\bf 0810}, 009 (2008)
  [arXiv:0808.2354 [hep-th]].

\bibitem{Hofman:2008ar} 
  D.~M.~Hofman and J.~Maldacena,
  ``Conformal collider physics: Energy and charge correlations,''
  JHEP {\bf 0805}, 012 (2008)
  [arXiv:0803.1467 [hep-th]].

\bibitem{Camanho:2009vw} 
  X.~O.~Camanho and J.~D.~Edelstein,
  ``Causality constraints in AdS/CFT from conformal collider physics and Gauss-Bonnet gravity,''
  JHEP {\bf 1004}, 007 (2010)
  [arXiv:0911.3160 [hep-th]].



\bibitem{Buchel:2009sk} 
  A.~Buchel, J.~Escobedo, R.~C.~Myers, M.~F.~Paulos, A.~Sinha and M.~Smolkin,
  ``Holographic GB gravity in arbitrary dimensions,''
  JHEP {\bf 1003}, 111 (2010)
  [arXiv:0911.4257 [hep-th]].



\end{thebibliography}
\end{document}